\documentclass[usenatbib]{mnras}
\usepackage{epsfig,booktabs,caption,fixltx2e,times,graphicx,amsmath}
\usepackage[flushleft]{threeparttable}

\def\Hy@FixNotFirstPage{%
	\gdef\Hy@FixNotFirstPage{%
		\setbox\AtBeginShipoutBox=\hbox{%
			\copy\AtBeginShipoutBox
		}%
	}%
}
\AtBeginShipout{\Hy@FixNotFirstPage}
 
\def\I{\,\textsc{i}}
\def\II{\,\textsc{ii}}

\def\IV{\,\textsc{iv}}

\def\hst{{\it HST}}

\def\spitzer{{\it Spitzer}}

\title[Progenitor and outburst of Gaia16cfr]{Connecting the progenitors, pre-explosion variability, and giant outbursts of luminous blue variables with Gaia16cfr}

\author[Kilpatrick et al.]{Charles D. Kilpatrick$^1$\thanks{Email:
    cdkilpat@ucsc.edu}, Ryan J. Foley$^1$, Maria R. Drout$^2$, Yen-Chen Pan$^1$,
	\newauthor Fiona H. Panther$^{3,4}$, David A. Coulter$^1$, Alexei V. Filippenko$^{5,6}$, G. H. Marion$^7$,
	\newauthor Anthony L. Piro$^2$, Armin Rest$^8$, Ivo R. Seitenzahl$^{9,3}$, Giovanni Strampelli$^8$, Xi E. Wang$^3$ \\
	$^1$Department of Astronomy and Astrophysics, University of California, Santa Cruz, CA 95064, USA\\
	$^2$Carnegie Observatories, 813 Santa Barbara Street, Pasadena, CA 91101, USA\\
	$^3$Research School of Astronomy and Astrophysics, Australian National University, Canberra, Australia\\
	$^4$ARC Centre of Excellence for All-Sky Astrophysics (CAASTRO), Canberra 2611, Australia\\
	$^5$Department of Astronomy, University of California, Berkeley, CA 94720-3411, USA\\
	$^6$Senior Miller Fellow, Miller Institute for Basic Research in Science, University of California, Berkeley, CA 94720, USA\\
	$^7$Department of Astronomy, University of Texas at Austin, 1 University Station C1400, Austin, TX, 78712-0259, USA\\
	$^8$Space Telescope Science Institute, 3700 San Martin Drive, Baltimore, MD 21218, USA\\
	$^9$School of Physical, Environmental, and Mathematical Sciences, University of New South Wales, Australian Defense Force Academy, Canberra, ACT 2600, Australia}

\begin{document}
\date{Accepted 0000, Received 0000, in original form 0000}
\pagerange{\pageref{firstpage}--\pageref{lastpage}} \pubyear{2017}
\maketitle
\label{firstpage}

\begin{abstract}
\noindent

We present multi-epoch, multi-colour pre-outburst photometry and post-outburst light curves and spectra of the luminous blue variable (LBV) outburst Gaia16cfr discovered by the {\it Gaia} satellite on 1 December 2016 UT.  We detect Gaia16cfr in 13 epochs of {\it Hubble Space Telescope} imaging spanning phases of 10~yr to 8 months before the outburst and in {\it Spitzer Space Telescope} imaging 13~yr before outburst.  Pre-outburst optical photometry is consistent with an 18~M$_{\odot}$ F8~I star, although the star was likely reddened and closer to 30~M$_{\odot}$. The pre-outburst source exhibited a significant near-infrared excess consistent with a $120$~AU shell with $4\times10^{-6}$~M$_{\odot}$ of dust. We infer that the source was enshrouded by an optically-thick and compact shell of circumstellar material from an LBV wind, which formed a pseudo-photosphere consistent with S~Dor-like variables in their ``maximum'' phase. Within a year of outburst, the source was highly variable on $10$--$30$~day timescales. The outburst light curve closely matches that of the 2012 outburst of SN~2009ip, although the observed velocities are significantly slower than in that event.  In H$\alpha$, the outburst had an excess of blueshifted emission at late times centred around $-1500~\text{km~s}^{-1}$, similar to that of double-peaked Type IIn supernovae and the LBV outburst SN~2015bh.  From the pre-outburst and post-outburst photometry, we infer that the outburst ejecta are evolving into a dense, highly structured circumstellar environment from precursor outbursts within years of the December 2016 event.

\end{abstract}


\section{INTRODUCTION}\label{sec:introduction}

There is an increasing sample of luminous transients associated with outbursts from $\ga20$~M$_{\odot}$ stars.  These events are usually less luminous than bona fide supernovae (SNe; with $M>-14$~mag) and exhibit Balmer lines with a full width at half-maximum intensity (FWHM) of $100$--$500~\text{km~s}^{-1}$).  Because of their spectroscopic characteristics and low luminosities, these outbursts are often confused with Type IIn SNe \citep[SNe~IIn; SNe defined by relatively narrow ($\la1000~\text{km~s}^{-1}$) lines of hydrogen in their spectra; see][]{schlegel90,filippenko97}, earning them the label ``SN impostors''\footnote{Although ``SN impostor'' implies that these objects are not genuine core-collapse SNe and thus confuses a physical mechanism with an observational class of transients, we use this label for consistency with the existing literature.  This does not imply that we think a single physical mechanism powers these objects or that none of these objects is a core-collapse SN.}. Many of these SN impostors have pre-outburst detections, which suggests their progenitor systems contain very massive stars. Given the luminosity, colours, and pre-outburst variability of these sources, several historical and recent SN impostors are thought to come from luminous blue variable (LBV) stars, including SN~1954J \citep{tammann+68}, SN~1997bs \citep{van-dyk+00}, SN~2000ch \citep{wagner+04,pastorello+10}, SN~2002kg \citep{weis+05,maund+06}, SN~2008S \citep{smith+09}, SN~2009ip and UGC2773-OT \citep{smith+10,foley+11}, SN~2015bh \citep[also known as SNHunt~275 and PTF13efv;][]{ofek+16,elias-rosa+16,thone+17}, and PSN J09132750+7627410 \citep{tartaglia+16}.  Light echoes from the Galactic LBV $\eta$~Car indicate that many SN impostors are spectroscopically similar to the nonterminal great eruption of that star in the 1830s, which ejected a massive, bipolar nebula of circumstellar material (CSM) but left a surviving star \citep{prieto+14}.

However, this interpretation may not hold true for some or all SN impostors.  \citet{prieto+08} found that the pre-outburst \spitzer\ luminosity of SN~2008S was consistent with a $10$~M$_{\odot}$ star, which suggests that the progenitor was an asymptotic giant branch (AGB) star or a red supergiant \citep[also][]{botticella+09}.  Extreme AGB stars are expected to be heavily obscured by dusty shells, which suggests they may be prime candidates for some SN~IIn progenitor systems, but also that precise identification of their progenitor systems may be difficult \citep{thompson+09,kochanek+11}.  Two or more populations of SN impostor progenitor systems may exist \citep{kochanek+12}, where a population of $<25~{\rm M}_{\odot}$ stars dominates ``SN 2008S-like'' transients, but any surviving star is obscured by dust reforming in the post-shock circumstellar environment.  These events contrast with extremely luminous outbursts such as SN~1961V, which appear to require higher mass ($>40~{\rm M}_{\odot}$), $\eta$ Car-like stars \citep{kochanek+12}.

It has also been hypothesised that some low-luminosity SNe~II-P (SNe that whose light curves ``plateau'' after peak luminosity, consistent with recombination of an extended stellar envelope or circumstellar, hydrogen-rich shell) are in fact SN impostors, and the progenitors of these events could also be relatively low-mass red supergiants \citep{dessart+10}.  The origin of these events and their association with progenitor systems having a wide mass range suggest that we must draw a connection between pre-outburst and post-outburst properties in order to fully understand the physical mechanism behind the outburst itself.

The origin of this mechanism is still unclear, especially as it must provide enough energy to the outburst without completely disrupting the progenitor star.  Galactic LBVs are usually defined by their characteristic S~Dor-like variability \citep[named for the prototypical LBV S~Doradus;][]{hubble+53,sharov+75,wolf+80} --- that is, variability in optical bands at roughly constant luminosity \citep{wolf+86,lamers+86,humphreys+88,wolf+89}.  However, the cycles of LBV variability from their ``minimum'' or hot, ultraviolet (UV) bright, quiescent phase to ``maximum'' or cool, optically bright, outburst phase occur over years or decades, likely from the formation of a dense, optically-thick wind during optical maximum that increases their apparent photospheric radii \citep[][]{massey+00,van-genderen+01}.  Many (although not all) LBVs also exhibit signatures of recent $\eta$~Car-like outbursts in the form of massive, bipolar nebulae \citep[e.g., AG~Car, HR~Car, HD~168625, He~3-519, P~Cygni, Sher~25, WRA~751;][]{johnson+76,johnson+92,smith+94,hutsemekers+94a,hutsemekers+94b,weis+97,weis+00,pasquali+02,groh+06,groh+09,weis+11}, which suggests they underwent relatively rapid changes in luminosity on short (month to year) timescales. The connection between these types of variability and their underlying physical mechanisms is still ambiguous, especially in LBVs where both are thought to occur.  S~Dor and $\eta$~Car-like variability appear to require periods of enhanced mass loss, but the magnitude, frequency, and duration of this mass loss can vary significantly \citep[see, e.g., the review by][]{vink11}.  

Curiously, while $\eta$~Car clearly survived its great eruption, it remains possible that some transients identified as SN impostors require more energy than an LBV outburst can provide and are actually core-collapse SNe. SN~1961V in NGC~1058 was historically interpreted as an LBV eruption given its low ejecta velocities and peculiar variability after peak luminosity \citep{humphreys+99}, but has since been reinterpreted as a core-collapse SN \citep[as originally proposed by][]{zwicky+64}. This interpretation is supported by the fact that SN~1961V was luminous for a SN impostor at peak ($M\approx -18$~mag), as well as {\it Spitzer} imaging, which placed deep upper limits below the level expected for any surviving star \citep{kochanek+11,smith+11}.  Subsequent analysis of the SN~1961V site using the {\it Hubble Space Telescope} (\hst) suggests this interpretation may be incorrect; there is a source consistent with a quiescent LBV at the site of the explosion \citep{van-dyk+12}. This type of analysis is complicated by the presence of dust formed in the ejecta and the origin of any infrared (IR) excess (or lackthereof) in the overall spectral energy distribution (SED). A star may have survived the outburst, but high extinction can obscure most of the UV/optical emission from any surviving star.  Deep late-time imaging of SN impostors is therefore critical, and studies of SN~1997bs \citep{adams+15}, SN~2008S and NGC 300-OT \citep{adams+16}, and SNHunt~248 \citep{mauerhan+17} indicate that some events fade well below the luminosity of their progenitor stars in the optical and near-infrared. However, this type of analysis can take years before emission from the outburst has faded to a level where the presence of a surviving star can be satisfactorily ruled out.

SN~2009ip was also identified as an LBV outburst \citep{smith+10,foley+11} with a subsequent transient from the same source in 2012 that may have been a core-collapse SN from the same star \citep{mauerhan+13}. The high peak luminosity ($M\approx -18$~mag) of the transient and broad spectral features (FWHM = $8000$~km s$^{-1}$) were interpreted as the signature of a core-collapse SN.  Other studies examining the 2012 outburst suggest that the limited energy in the outburst may be inconsistent with a core-collapse SN \citep{margutti+14}.  Spectropolarimetry of SN~2009ip suggests that the low apparent energy may be the consequence of a toroidal distribution of CSM around the explosion; only a small fraction of the outburst ejecta interacted with CSM to produce radiation \citep{mauerhan+14}.  In this way, the total energy of the outburst could be much closer to $10^{51}$~erg as expected for a core-collapse SN.  However, the lack of any nebular features even at extremely late times ($>1000$~days) after peak luminosity is inconsistent with most models of core-collapse SNe \citep[][although the exact nature of the circumstellar interaction complicates this interpretation]{fraser+13,graham+17}.  Alternative explanations for SN~2009ip include a nonterminal pulsational pair instability SN, especially considering the high inferred mass of the progenitor star \citep[$50$--$80$~M$_{\odot}$][]{smith+10,fraser+13,woosley+17}, although this model does not accurately predict the timescale of pulses and ejecta mass of SN~2009ip or rates for SN~2009ip-like events \citep{ofek+13,smith+14}.

In this paper, we discuss the massive-star outburst Gaia16cfr\footnote{This name was adopted from \citet{atel9937} and subsequent Astronomer's Telegrams.} in NGC~2442.  Gaia16cfr was discovered at $\alpha=7^{\text{h}}36^{\text{m}}25^{\text{s}}.96$, $\delta=-69^{\circ}32\arcmin55\arcsec.26$ by the {\it Gaia} satellite on 1 December 2016 (UT dates are used throughout this paper) with $G=19.3$~mag\footnote{\url{http://gsaweb.ast.cam.ac.uk/alerts/alert/Gaia16cfr/.}}, corresponding to an absolute magnitude of $M_G=-12.2$~mag.  Given this low luminosity and the presence of narrow P-Cygni Balmer lines in follow-up spectra, \citet{atel9937} identified Gaia16cfr as a likely SN impostor.  NGC~2442 was the host of the peculiar low-luminosity SN~II 1999ga \citep{pastorello+09} as well as the SN~Ia 2015F and had been observed with deep, multi-band, multi-epoch \hst\ imaging by \citet{riess+16}, who derived a Cepheid distance modulus of $m-M=31.51\pm0.05$~mag ($20.1\pm0.5$~Mpc). \citet{atel9938} and \citet{atel9982} identified a counterpart to Gaia16cfr in pre-outburst \hst\ images. The luminosity of this counterpart was consistent with a relatively low-mass ($<20$~M$_{\odot}$) source, but also one that was highly variable and significantly reddened within a year of outburst. 

Here, we present the entire pre-outburst \hst\ light curve of Gaia16cfr, as well as detections of a potential counterpart in pre-outburst \spitzer/IRAC imaging and post-outburst photometry and spectroscopy.  We analyse the full SED of the pre-outburst photometry, which demonstrates that the source was in a dusty environment and is consistent with a $>18$~M$_{\odot}$ star.  Variability in the pre-outburst light curve of Gaia16cfr is similar to the ``flickering'' observed in pre-outburst light curves of other SN impostors such as SN~1954J and SN~2009ip \citep{tammann+68,smith+10}. We demonstrate that the outburst light curve is consistent with that of the highest luminosity outbursts, such as SN~1961V, SN~2015bh, and the 2012 outburst of SN~2009ip \citep[which we refer to as SN~2009ip-12B following the convention of][where SN~2009ip-12A refers to one of the precursor outbursts that occurred within $\sim40$~days of the rise to peak]{pastorello+13,graham+14}.  From spectroscopy and photometry of Gaia16cfr, we find that the apparent blackbody temperature of the continuum emission cooled rapidly within $120$~days of discovery. The H$\alpha$ emission line exhibited a double-peaked profile with significant blueshifted excess, which we interpret as an interaction between an ejecta shell and previously ejected CSM that is becoming optically thin.  We discuss the structure of the circumstellar environment around Gaia16cfr in light of these findings, as well as the mass-loss history of its progenitor star. Throughout this paper, we assume the above Cepheid distance to NGC~2442 ($20.1\pm0.5$~Mpc) and a Milky Way extinction of $A_{V} = 0.556$~mag \citep{schlafly+11}.   

\section{OBSERVATIONS}\label{sec:observations}

\subsection{Archival Data}

\subsubsection{{\it Hubble Space Telescope}}

We obtained \hst/ACS imaging of NGC~2442 from the \hst\ Legacy Archive\footnote{\url{https://hla.stsci.edu/hla_faq.html}} from 20 Oct. 2006 (Cycle 15, Program GO-10803, PI Smartt) as well as \hst/WFC3 imaging from 21 Jan. 2016 to 9 Apr. 2016 (Cycle 22, Program GO-13646, PI Foley). These data were processed using the latest calibration software and reference files, which included corrections for bias, dark current, flat-fielding, and bad-pixel masking. Where there were multiple exposures per epoch, individual frames were processed and combined using the {\tt IRAF}\footnote{IRAF, the Image Reduction and Analysis Facility, is distributed by the National Optical Astronomy Observatory, which is operated by the Association of Universities for Research in Astronomy (AURA) under cooperative agreement with the National Science Foundation (NSF).} task {\tt MultiDrizzle}, which performs registration, cosmic-ray rejection, and final image combination using the {\tt Drizzle} task. We performed photometry on these combined images in each filter using the {\tt dolphot}\footnote{\url{http://americano.dolphinsim.com/dolphot/}} stellar photometry package to obtain instrumental magnitudes for sources in each image.  We calibrated these instrumental magnitudes using zeropoints from the ACS/WFC zeropoint calculator tool for 20 Oct. 2006\footnote{\url{https://acszeropoints.stsci.edu/}} and from the most up-to-date  WFC3/UVIS and WFC3/IR photometric zeropoints available at \url{http://www.stsci.edu/hst/wfc3/analysis/}.

\begin{table}
\begin{center}\begin{minipage}{3.3in}
      \caption{Swope Optical Photometry of Gaia16cfr}\scriptsize
  	\setlength\tabcolsep{2.5pt}
\begin{tabular}{@{}lcccccc}\hline\hline
 Julian Date & $u$ & $B$ & $V$ & $g$ & $r$ & $i$ \\ \hline
2457774.85 & 14.823 (161) & 14.740 (003) & 14.489 (004) & 14.543 (002) & 14.546 (003) & 14.621 (008) \\
2457780.71 & 14.598 (160) & 14.405 (003) & 14.104 (004) & 14.199 (003) & 14.170 (003) & 14.161 (005) \\
2457781.80 & 14.558 (164) & 14.402 (003) & 14.087 (003) & 14.175 (003) & 14.129 (035) & 14.128 (005) \\
2457782.75 & 14.572 (172) & 14.381 (004) & 14.059 (003) & 14.181 (003) & 14.128 (003) & 14.127 (004) \\
2457784.78 & 14.656 (192) & 14.254 (014) & 13.882 (014) & 14.170 (004) & 14.112 (004) & 14.136 (011) \\
2457792.79 & 15.866 (182) & 15.116 (004) & 14.646 (004) & 14.821 (003) & 14.604 (005) & 14.542 (014) \\
2457801.71 & 16.749 (538) & 16.101 (006) & 15.341 (005) & 15.627 (004) & 15.266 (005) & 15.247 (010) \\
2457803.69 & 17.277 (229) & 16.258 (005) & 15.563 (006) & 15.856 (005) & 15.402 (004) & 15.442 (006) \\
2457806.71 & --- & 16.540 (005) & 15.755 (006) & 16.053 (004) & 15.592 (004) & 15.539 (005) \\
2457808.65 & 17.729 (173) & 16.653 (005) & 15.848 (005) & 16.168 (005) & 15.669 (004) & 15.600 (006) \\
2457816.69 & 18.275 (163) & 17.019 (006) & 16.159 (008) & 16.550 (006) & 15.945 (004) & 15.882 (011) \\
2457818.60 & --- & --- & --- & 16.617 (008) & 16.007 (006) & 15.933 (007) \\
2457821.66 & --- & --- & --- & 16.733 (009) & 16.096 (005) & 16.039 (008) \\
2457823.63 & --- & --- & --- & 16.851 (010) & 16.167 (007) & 16.099 (010) \\
2457826.63 & --- & --- & --- & 16.992 (011) & 16.291 (007) & 16.163 (008) \\
2457828.62 & --- & --- & --- & 17.073 (009) & 16.311 (007) & 16.237 (009) \\
2457831.66 & --- & --- & --- & 17.207 (009) & 16.409 (008) & 16.314 (011) \\
2457832.55 & --- & --- & --- & 17.323 (007) & 16.474 (006) & 16.403 (012) \\
2457833.56 & --- & --- & --- & 17.410 (064) & 16.576 (048) & 16.390 (050) \\
2457849.65 & --- & --- & --- & 18.585 (116) & 17.794 (085) & 17.723 (093) \\
2457852.67 & --- & --- & --- & 18.494 (115) & 17.777 (087) & 17.744 (093) \\
2457864.60 & --- & --- & --- & 18.847 (130) & 18.085 (099) & 18.041 (114) \\
2457869.57 & --- & --- & --- & 18.869 (129) & 18.099 (099) & 18.096 (112) \\
2457871.56 & --- & --- & --- & 18.885 (130) & 18.121 (100) & 18.106 (111) \\
2457876.54 & --- & --- & --- & 19.000 (136) & 18.221 (103) & 18.319 (118) \\
2457882.55 & --- & --- & --- & 18.938 (146) & 18.313 (114) & 18.391 (125) \\
2457888.59 & --- & --- & --- & 19.043 (158) & 18.343 (118) & 18.402 (133) \\
2457893.51 & --- & --- & --- & 19.145 (150) & 18.316 (110) & 18.440 (130) \\
2457907.51 & --- & --- & --- & 19.401 (191) & 18.388 (162) & 18.619 (178) \\
2457909.49 & --- & --- & --- & 19.410 (170) & 18.516 (121) & 18.677 (145) \\
2457915.52 & --- & --- & --- & 19.402 (176) & 18.621 (121) & 18.729 (145) \\
\hline
\end{tabular}\label{tab:phot}
\begin{tablenotes}
      \small 
\item {\bf Note.} Uncertainties ($1\sigma$) are in millimagnitudes and given in parentheses next to each measurement. $BV$ magnitudes are on the Vega scale and $ugri$ are on the AB scale.
\end{tablenotes}
\end{minipage}
\end{center}
\end{table}

\subsubsection{{\it Spitzer Space Telescope}/IRAC}

We obtained a 30~s \spitzer/IRAC exposure of NGC~2442 taken on 21 Nov. 2003 from the \spitzer\ Heritage Archive (AOR-7858176, PI Fazio).  The pipeline-reduced and calibrated images were processed using {\tt MOPEX}, and each channel was combined into a single frame with a scale of 0\arcsec.6 pixel$^{-1}$.  Although the pre-outburst source may have been relatively bright in IRAC bands owing to dust emission, the source was in a crowded field and close to the southern spiral arm of NGC~2442.  Therefore, we used the {\tt IRAF} task {\tt daophot} with a point-spread function (PSF) constructed empirically from bright field stars well-separated from the centre of NGC~2442. We used this PSF to perform unforced photometry of all point sources in each of the IRAC frames and estimate the Poisson and background noise associated with each source.  Each measurement was calibrated using photometric zeropoints given in the IRAC instrument handbook for the cold \spitzer\ mission\footnote{\url{http://irsa.ipac.caltech.edu/data/SPITZER/docs/irac/iracinstrumenthandbook/17/}}. 

\begin{figure*}
	\includegraphics[width=0.33\textwidth]{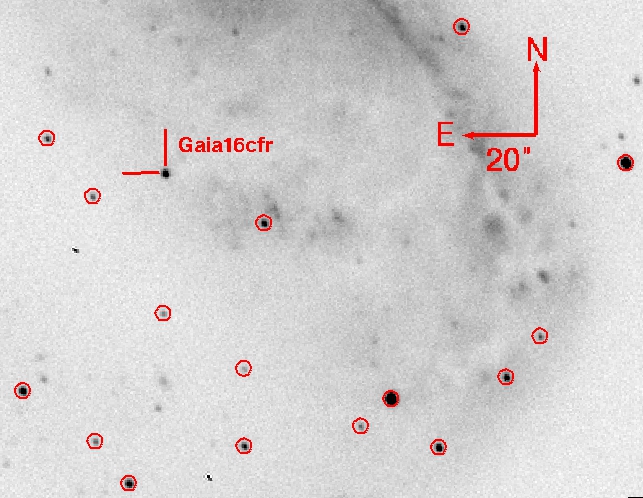}
	\includegraphics[width=0.33\textwidth]{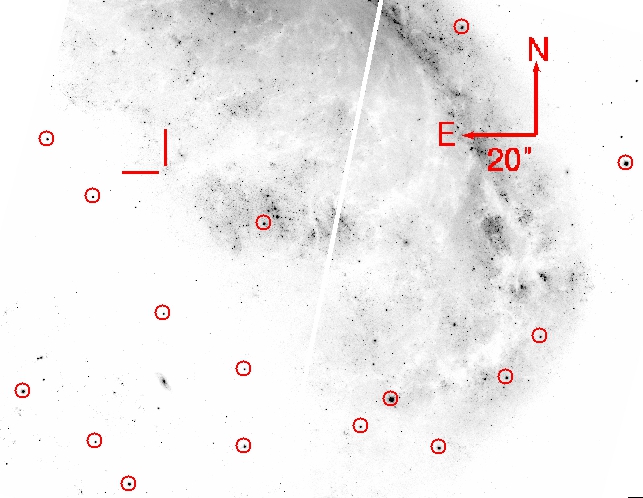}
	\includegraphics[width=0.33\textwidth]{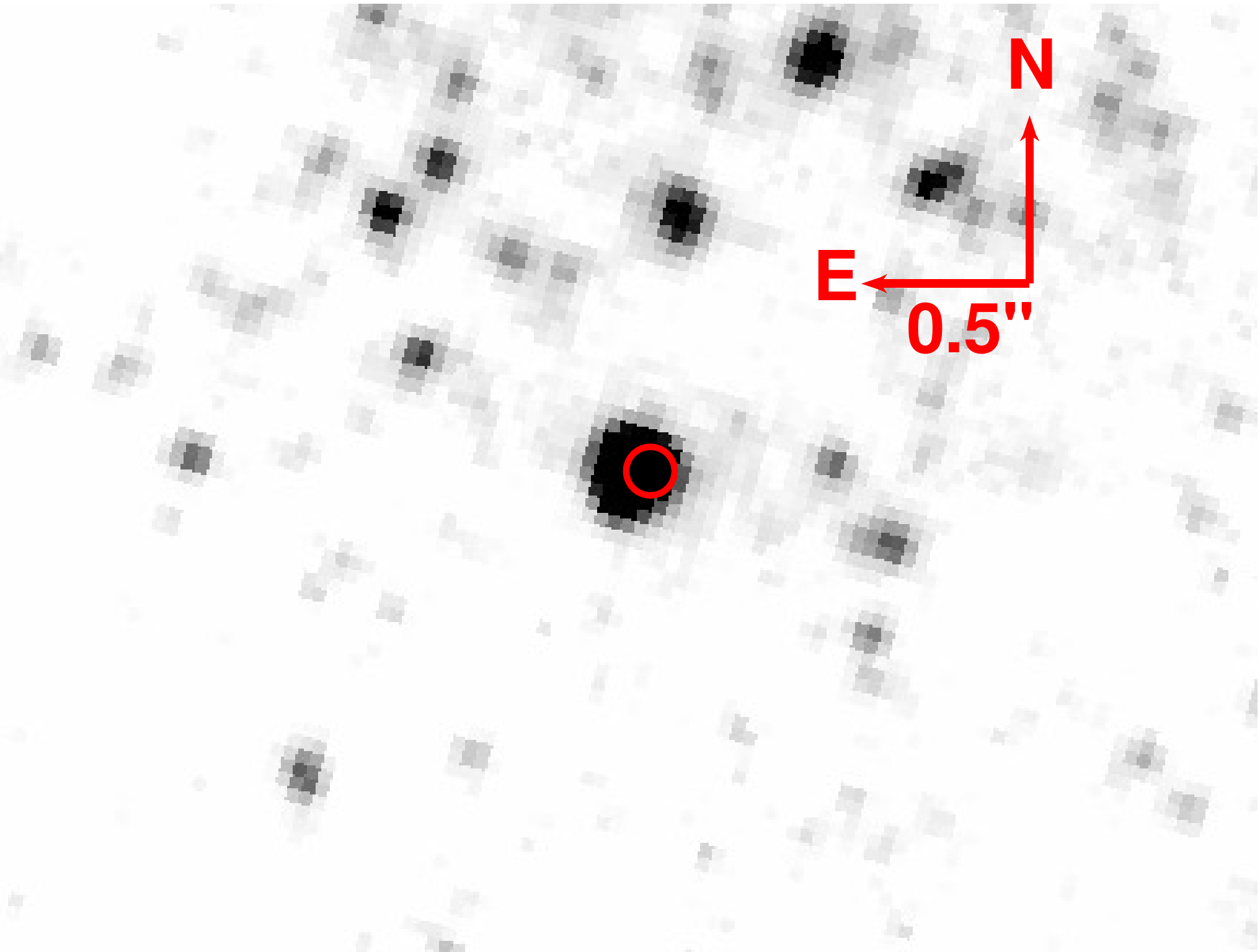}
	\caption{(Left) A $170\arcsec \times 130\arcsec$ cutout of our Swope $r$-band image of NGC~2442 from 6 Apr. 2017 with the position of Gaia16cfr marked and 16 point sources used for astrometry circled.  (Middle) \hst/WFC3 $F350LP$ image from 8 Feb. 2016 with the same 16 point sources used for astrometry circled and the position of Gaia16cfr marked.  (Right) A $3.7\arcsec \times 2.9\arcsec$ region of the middle panel centred on the location of Gaia16cfr.  The position of Gaia16cfr is marked with an ellipse having semimajor and semiminor axes $0.072\arcsec$ and $0.070\arcsec$, respectively (i.e., the approximate, combined astrometric uncertainty in the position of Gaia16cfr and relative astrometry in right ascension and declination).  The position of Gaia16cfr is consistent with a single point source to $1\sigma$ uncertainty and there are no other point sources within $\sim7.1\sigma$.}\label{fig:hst}
\end{figure*}

\subsubsection{ESO NTT + EFOSC2}

We obtained imaging of Gaia16cfr from the European Southern Observatory (ESO) public data archive which was taken as part of PESSTO \footnote{www.pessto.org} \citep{smartt15}.  The images were taken with the ESO Faint Object Spectrograph and Camera (EFOSC2) on the ESO 3.6~m New Technology Telescope (NTT) at La Silla Observatory, Chile. The data consisted of 8 frames taken with the Bessel $V$ filter between 4 Jan. and 7 Feb. 2017.  The exposure time varied from image to image, but was generally around $20$~s.  We performed PSF photometry on these images using the {\tt IRAF} task {\tt daophot} and calibrated the instrumental magnitudes using APASS $V$-band stars \citep{henden+16}. The magnitudes of Gaia16cfr are presented in \autoref{tab:efosc}.

\begin{table}
\begin{center}\begin{minipage}{3.3in}
      \caption{EFOSC2 Photometry}\scriptsize
  	\setlength\tabcolsep{3.0pt}
\begin{tabular}{@{}lcc}\hline\hline
  Julian Date  & $V$ ($\sigma$) (mag) \\ \hline
2457757.78 & 18.65 (040) \\
2457759.84 & 18.76 (040) \\
2457770.68 & 15.60 (090) \\
2457771.80 & 15.49 (080) \\
2457773.71 & 15.08 (070) \\
2457779.76 & 14.64 (050) \\
2457780.68 & 14.45 (110) \\
2457781.73 & 14.53 (070) \\
\hline
\end{tabular}\label{tab:efosc}
\end{minipage}
\end{center}
\end{table}

\subsection{Swope Photometry}

We observed Gaia16cfr using the Direct CCD Camera on the Swope 1.0~m telescope at Las Campanas Observatory, Chile, between 21 Jan. 2017 and 11 June 2017 in $uBVgri$\footnote{Swope filter functions are provided at \url{http://csp.obs.carnegiescience.edu/data/filters}}. Standard reductions were performed on these data using the {\tt photpipe} imaging and photometry package \citep{rest+05}. A robust pipeline used by several time-domain surveys \citep[e.g., Pan-STARRS1;][]{rest+14}, {\tt photpipe} is designed to perform single-epoch image processing, including image calibration (e.g., bias subtraction, cross-talk corrections, flat-fielding), astrometric calibration, and dewarping \citep[using SWarp;][]{bertin+02}. Unlike most {\tt photpipe} applications, we did not perform template subtraction. We used {\tt DoPhot} \citep{schechter+93} optimised for PSF photometry on the reduced images to obtain instrumental magnitudes of Gaia16cfr and nearby standard stars.  Finally, we calibrated our $uBVgri$ photometry using PS1 standard-star fields observed in the same instrumental configuration and at a similar airmass.  The PS1 magnitudes were transformed into the Swope natural system using Supercal transformations as described by \citet{scolnic+15}.  We verified our calibration using the same $BVgri$ photometric standards used for SN~2015F by \citet[][Table A1]{cartier+17}, which agree with our measurements to within the $1\sigma$ uncertainties.  Our $uBVgri$ photometry of Gaia16cfr is presented in \autoref{tab:phot}.
 
\subsection{Spectroscopy}

We observed Gaia16cfr on 19 Jan., 29 Mar., and 1 and 29 May 2017 with the Goodman Spectrograph \citep{clemens+04} on the 4.1~m Southern Astrophysical Research Telescope (SOAR) on Cerro Pach\'{o}n, Chile. The 1.07\arcsec\ slit was used in conjunction with the 400~l~mm$^{-1}$ grating for an effective spectral range of 4000--7050~\AA\ in our blue setup and 5000--9050~\AA\ in our red setup. We used a blocking filter (GG~455) in the red setup to minimise second-order blue-light contamination. The airmass was moderate (1.3--1.5) during most of our spectral epochs, so we aligned the slit to the parallactic angle to minimise the effects of atmospheric dispersion \citep{filippenko82}. Standard reductions of the two-dimensional (2D) spectra were performed using {\tt IRAF}. We used the {\tt IRAF} task {\tt apall} to optimally extract the one-dimensional (1D) blue and red spectra. Wavelength calibration on these one-dimensional images was done using calibration-lamp exposures taken immediately after each spectrum. Flux calibration was performed using a sensitivity function derived from standard-star spectra obtained on the same night and at similar airmass as each of our Gaia16cfr spectra. We dereddened each spectrum by $E(B-V)=0.18$~mag and removed the recession velocity $v = 1466$~km~s$^{-1}$, which is consistent with the velocity of the host galaxy. Finally, we combined the red and blue spectra into a single spectrum for each epoch.

We observed (PI Panther) Gaia16cfr on the night of 27 Jan. 2017 with the WiFeS Integral Field Spectrograph \citep{dopita+07} on the Australian National University 2.3~m telescope at Siding Springs Observatory for $2\times900$~s (coadded) under clear conditions and a typical seeing of 2\arcsec.  The observations were performed with the RT560 dichroic and the B3000 and R3000 gratings in place, giving a typical resolving power of $R=3000$. A 900~s sky exposure was used to remove night-sky lines, and the data were flux calibrated with the standard star HD~16031. The reduction, which includes dome and sky flat-fielding, wavelength calibration, bias subtraction, cosmic-ray rejection, atmospheric dispersion corrections, and telluric-line removal, was performed with the PyWiFeS pipeline \citep{childress+14}.

We obtained six spectra of Gaia16cfr with the Wide Field CCD (WFCCD) spectrograph mounted on the 2.5~m Ir\'{e}n\'{e}e du Pont Telescope at Las Campanas Observatory, spanning 2017 Jan. 25 to 2017 Mar. 28, and one spectrum with the Low Dispersion Survey Spectrograph 3 (LDSS-3) on the the 6.5~m Magellan/Clay telescope on 2017 Apr. 30.  WFCCD and LDSS spectra were obtained with the blue grism and VPH-All grism/blue-slit, respectively.  All spectra were observed with the slit aligned along the parallactic angle. 

Initial data reduction was performed using standard routines in {\tt IRAF}. 1D spectra were extracted using the {\tt IRAF} routine {\tt apall}, and wavelength calibration was performed using comparison-lamp exposures taken immediately after each science image. Flux calibration and telluric correction were performed using a set of custom {\tt idl} scripts \citep[see][]{matheson+08} and based on standard-star spectra obtained on the same night and at similar airmass to the spectra of Gaia16cfr.

We also obtained the classification spectrum of Gaia16cfr from the Transient Name Server\footnote{\url{https://wis-tns.weizmann.ac.il/object/2016jbu}}. This spectrum was taken by \citet{atel9938} within the PESSTO programme \citep{smartt15} on 2017 Jan. 3.  Our full spectroscopic series is summarised in \autoref{tab:spectroscopy}.

\begin{table}
\begin{center}\begin{minipage}{3.3in}
      \caption{Spectroscopy of Gaia16cfr}\scriptsize
  	\setlength\tabcolsep{3.0pt}
\begin{tabular}{@{}lcccc}\hline\hline
  Julian Date  & Telescope/Instrument & Range       & Grating/Grism & Exposure  \\
               &                      &  (\AA)      &               & (s)       \\ \hline
2457757.78     & NTT/EFOSC2           & 3638--9233  & Gr\#13        & 900        \\
2457772.70     & SOAR/Goodman         & 3600--9040  & R400/B400     & 1200/1200 \\
2457778.75     & du Pont/WFCCD WF4K–1 & 3702--9300  & Blue Grism    & $2\times600$     \\
2457780.94     & SSO/WiFeS            & 3500--9200  & B3000/R3000   & $2\times900$ \\
2457804.69     & du Pont/WFCCD WF4K–1 & 3702--9300  & Blue Grism    & $3\times300$     \\
2457808.65     & du Pont/WFCCD WF4K–1 & 3702--9300  & Blue Grism    & $3\times400$     \\
2457812.52     & du Pont/WFCCD WF4K–1 & 3702--9300  & Blue Grism    & $3\times300$     \\
2457838.61     & du Pont/WFCCD WF4K–1 & 3702--9300  & Blue Grism    & $2\times900$     \\
2457841.60     & SOAR/Goodman         & 3600--9020  & R400/B400     & 1200/1200 \\
2457840.73     & du Pont/WFCCD WF4K–1 & 3702--9300  & Blue Grism    & $2\times900$     \\
2457873.58     & Magellan/LDSS-3      & 4379--6506  & VPH-All grism & $2\times900$     \\
2457874.56     & SOAR/Goodman         & 3600--9000  & R400/B400     & 1500/1500 \\
2457902.57     & SOAR/Goodman         & 3600--9000  & R400/B400     & 1500/1500 \\
\hline
\end{tabular}\label{tab:spectroscopy}
\end{minipage}
\end{center}
\end{table}

\section{RESULTS}\label{sec:results}

\subsection{Astrometry Between Gaia16cfr and Pre-Outburst \hst\ Sources}\label{sec:astrom}

We used the {\tt photpipe} PSF-fit coordinates from our Swope $r$-band image of Gaia16cfr on 6 Apr. 2017 and the {\tt dolphot} PSF-fit coordinates of stars in the \hst\ frames to perform relative astrometry between these images.  For each \hst\ frame, we identified 12--16 sources common to both \hst\ and Swope imaging.  Using the coordinates of these sources, we calculated and applied a WCS solution for each \hst\ frame with the {\tt IRAF} tasks {\tt ccmap} and {\tt ccsetwcs}. We estimated the astrometric uncertainty of our geometric projection in the \hst\ image by selecting random subsamples consisting of half of our common stars, calculating a geometric projection, then determining the offsets between the remaining common stars. In this way, the astrometric uncertainty was generally $\sigma_{\alpha} = 0.05\arcsec$ (1.26 \hst/WFC3 pixels) and $\sigma_{\delta} = 0.038\arcsec$ (0.95 \hst/WFC3 pixels). On 6 Apr. 2017, Gaia16cfr was detected with the Swope telescope in $r$ with a signal-to-noise ratio (S/N) of $106$ at $\alpha= 7^{\text{h}}36^{\text{m}}25^{\text{s}}.965\pm0.011$, $\delta=-69^{\circ}32\arcmin55\arcsec.558\pm0.067$. At every \hst\ epoch, this position corresponds to a single point source to within $1\sigma$ astrometric precision. In \autoref{fig:hst}, we show our 6 Apr. 2017 Swope $r$-band image and 8 Feb. 2016 \hst\ WFC3/UVIS $F350LP$ image with $16$ common source circled. In the Swope image, we denote the position of Gaia16cfr. We also show a $3.7\arcsec \times 2.9\arcsec$ cutout of the same \hst\ image with the position of Gaia16cfr marked.  In this \hst\ epoch, there are no other point sources within $7.1\sigma$ of the Swope position. In \autoref{tab:hst}, we give our \hst\ magnitudes (in Vega magnitudes) for the source coincident with Gaia16cfr.

In all of our \hst\ imaging, the point source associated with Gaia16cfr is consistent with a single, unblended source.  There is effectively zero crowding around this source, indicating that it is likely an isolated, bright star.  The PSF of the source is similar to that of other isolated stars, with no indication of extended emission.  The {\tt dolphot} sharpness and roundness parameters were typically $-0.01$ to $-0.07$ and $0.02$ to $0.057$ (respectively), consistent with a single point source.  

We estimate the probability of a chance coincidence in the \hst\ images by noting that there are roughly $600$--$1300$ point sources with S/N $>3$ within a 20\arcsec\ radius of Gaia16cfr in each \hst\ image.  The 3$\sigma$ uncertainty ellipse for the \hst\ reference image has a solid angle of $\sim 0.18~\text{arcsec}^{2}$, which implies that 108--234~$\text{arcsec}^{2}$ or 8--19\% of the \hst\ image within $20\arcsec$ of the identified source has a point source that is close enough to be associated with that region. This value represents the probability that the detected point source is a chance coincidence. Thus, although it is unlikely that the identified point source is a chance coincidence, there is some probability that this is the case.  Follow-up imaging will be critical in order to confirm or rule out this possibility.

Gaia16cfr was also observed on 1 Feb. 2017 with \hst/WFC3 in $F814W$ in $6\times120$~s exposures (Cycle 24, Program GO-14645, PI Van Dyk).  We obtained this imaging from the MAST data archive\footnote{\url{https://archive.stsci.edu/}}, reduced each frame following standard image-reduction procedures, and then combined the individual frames with {\tt MultiDrizzle}.  Relative astrometry was performed between the combined frame and the pre-outburst \hst\ frames.  The location of Gaia16cfr was consistent with that of our Swope photometry and agrees with the same single point source in every pre-outburst image to within $1\sigma$ astrometric precision.

\begin{table}
\begin{center}\begin{minipage}{3.3in}
      \caption{\hst\ Photometry of the Gaia16cfr Counterpart}\scriptsize
\begin{tabular}{@{}cccccc}\hline\hline
  Julian Date  & Instrument & Filter   & Exp. Time (s)   & Magnitude ($1\sigma$) \\
  2454029.30   & ACS/WFC    & $F435W$  & $4\times395$    & 25.066 (025) \\
  2454029.37   & ACS/WFC    & $F658N$  & $3\times450$    & 21.193 (014) \\
  2454029.39   & ACS/WFC    & $F814W$  & $3\times400$    & 23.494 (016) \\
  2457408.60   & WFC3/UVIS  & $F350LP$ & $3\times420$    & 23.112 (006) \\
  2457408.67   & WFC3/IR    & $F160W$  & $2\times502.94$ & 20.280 (005) \\
  2457418.83   & WFC3/UVIS  & $F350LP$ & $3\times420$    & 21.752 (003) \\
  2457418.88   & WFC3/UVIS  & $F555W$  & $2\times488$    & 22.954 (011) \\
  2457427.45   & WFC3/UVIS  & $F350LP$ & $3\times420$    & 21.775 (003) \\
  2457427.47   & WFC3/IR    & $F160W$  & $2\times502.94$ & 19.160 (003) \\
  2457436.38   & WFC3/UVIS  & $F350LP$ & $3\times420$    & 22.742 (005) \\
  2457436.38   & WFC3/UVIS  & $F814W$  & $2\times488$    & 23.006 (017) \\
  2457442.02   & WFC3/UVIS  & $F350LP$ & $3\times420$    & 22.826 (005) \\
  2457442.08   & WFC3/IR    & $F160W$  & $2\times502.94$ & 20.309 (004) \\
  2457447.06   & WFC3/UVIS  & $F350LP$ & $3\times420$    & 23.348 (007) \\
  2457447.11   & WFC3/UVIS  & $F555W$  & $2\times488$    & 24.444 (021) \\
  2457451.51   & WFC3/UVIS  & $F350LP$ & $3\times420$    & 22.503 (004) \\
  2457451.61   & WFC3/IR    & $F160W$  & $2\times502.94$ & 19.939 (004) \\
  2457458.13   & WFC3/UVIS  & $F350LP$ & $3\times420$    & 22.321 (004) \\
  2457458.19   & WFC3/UVIS  & $F814W$  & $2\times488$    & 22.979 (018) \\
  2457463.16   & WFC3/UVIS  & $F350LP$ & $3\times420$    & 22.639 (005) \\
  2457463.23   & WFC3/IR    & $F160W$  & $2\times502.94$ & 20.004 (004) \\
  2457468.89   & WFC3/UVIS  & $F350LP$ & $3\times420$    & 22.934 (006) \\
  2457468.91   & WFC3/UVIS  & $F555W$  & $2\times488$    & 23.983 (017) \\
  2457477.83   & WFC3/UVIS  & $F350LP$ & $3\times420$    & 23.354 (008) \\
  2457477.85   & WFC3/IR    & $F160W$  & $2\times502.94$ & 20.900 (007) \\
  2457488.47   & WFC3/UVIS  & $F350LP$ & $3\times420$    & 23.600 (008) \\
  2457488.49   & WFC3/UVIS  & $F814W$  & $2\times488$    & 23.551 (023) \\
\hline
\end{tabular}\label{tab:hst}
\end{minipage}
\end{center}
\end{table}

\subsection{Pre-Outburst \spitzer\ Sources}\label{sec:astrom}

\begin{figure}
	\includegraphics[width=0.24\textwidth]{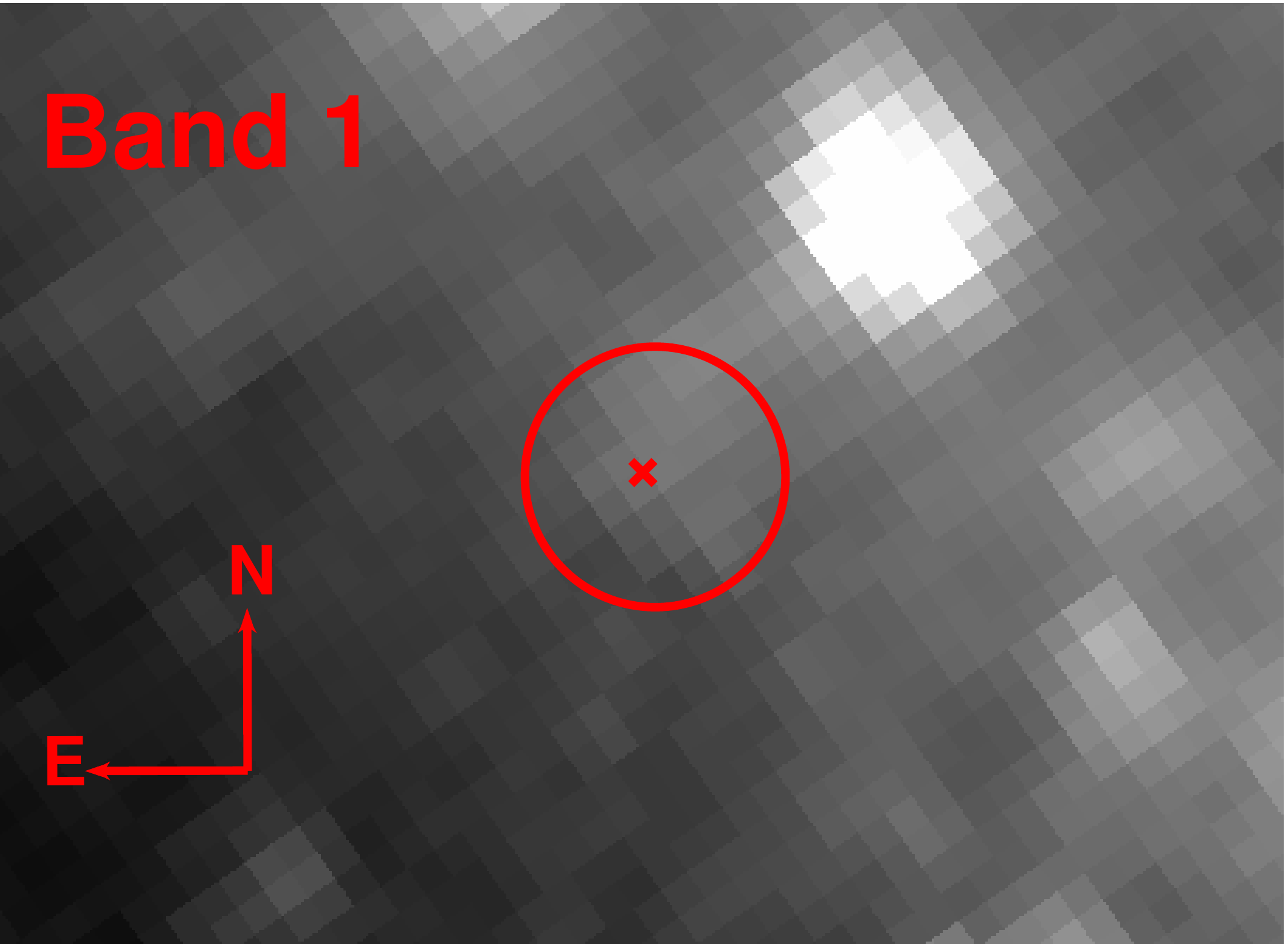}
	\includegraphics[width=0.24\textwidth]{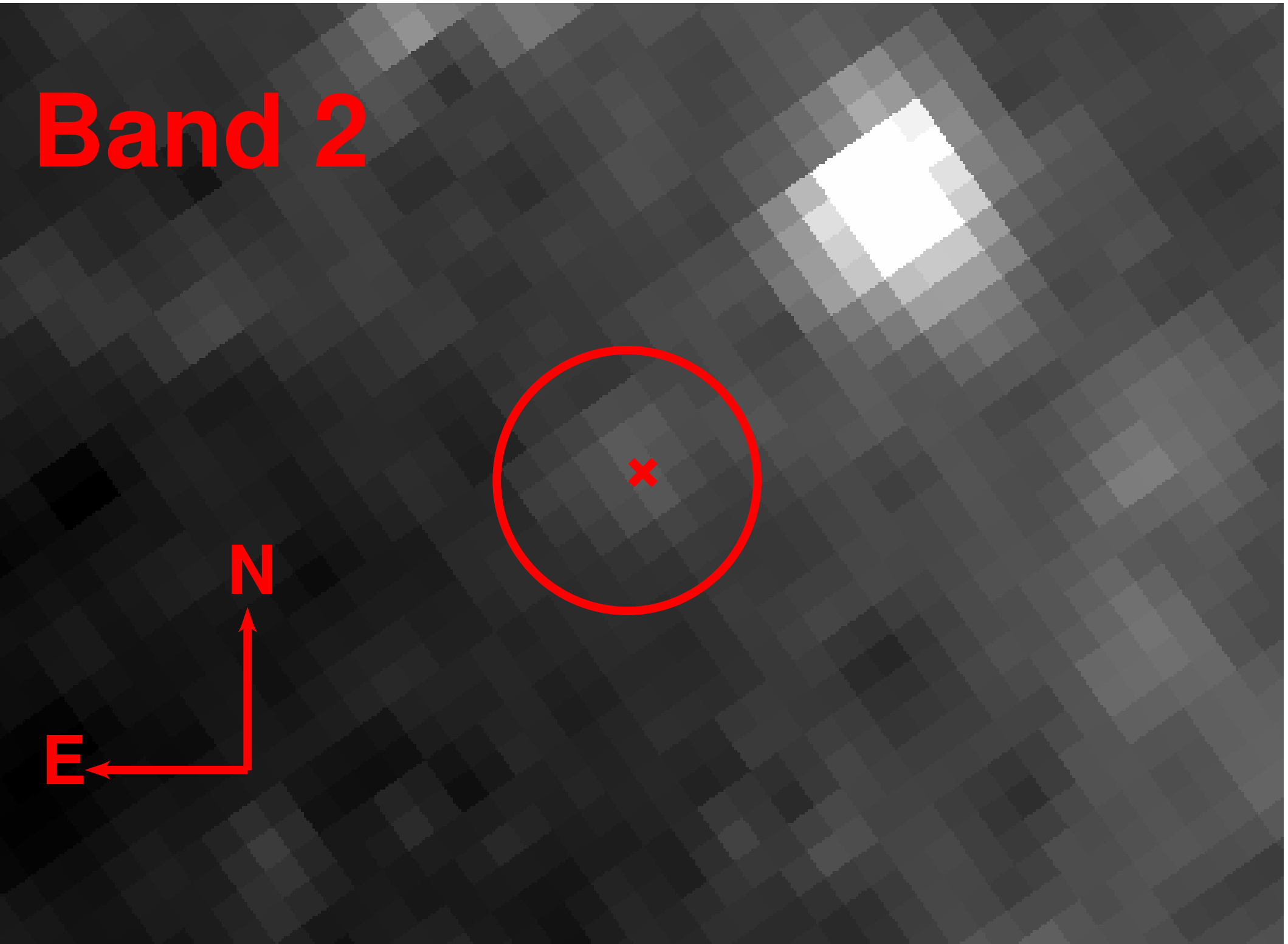}
	\includegraphics[width=0.24\textwidth]{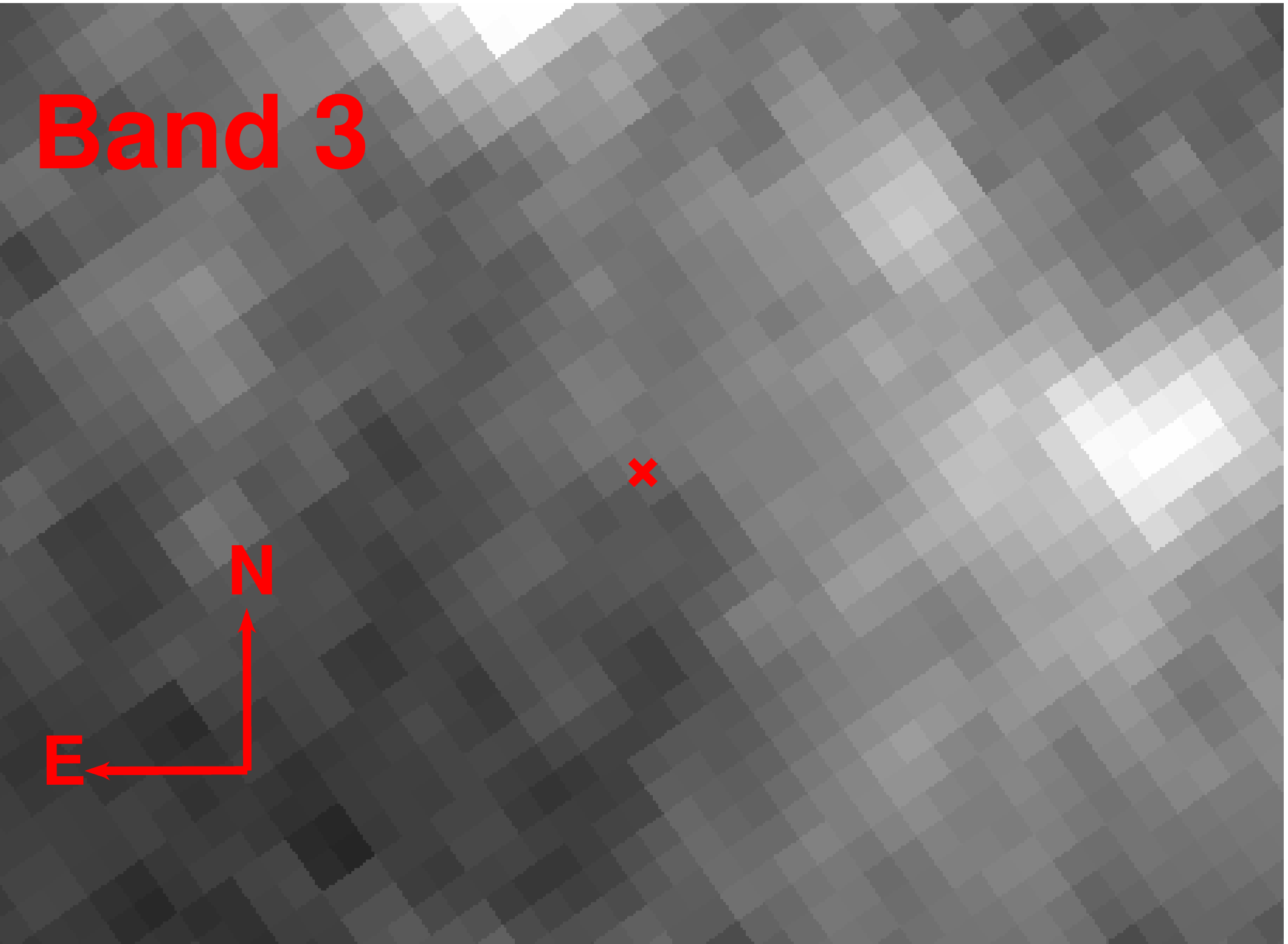}
	\includegraphics[width=0.24\textwidth]{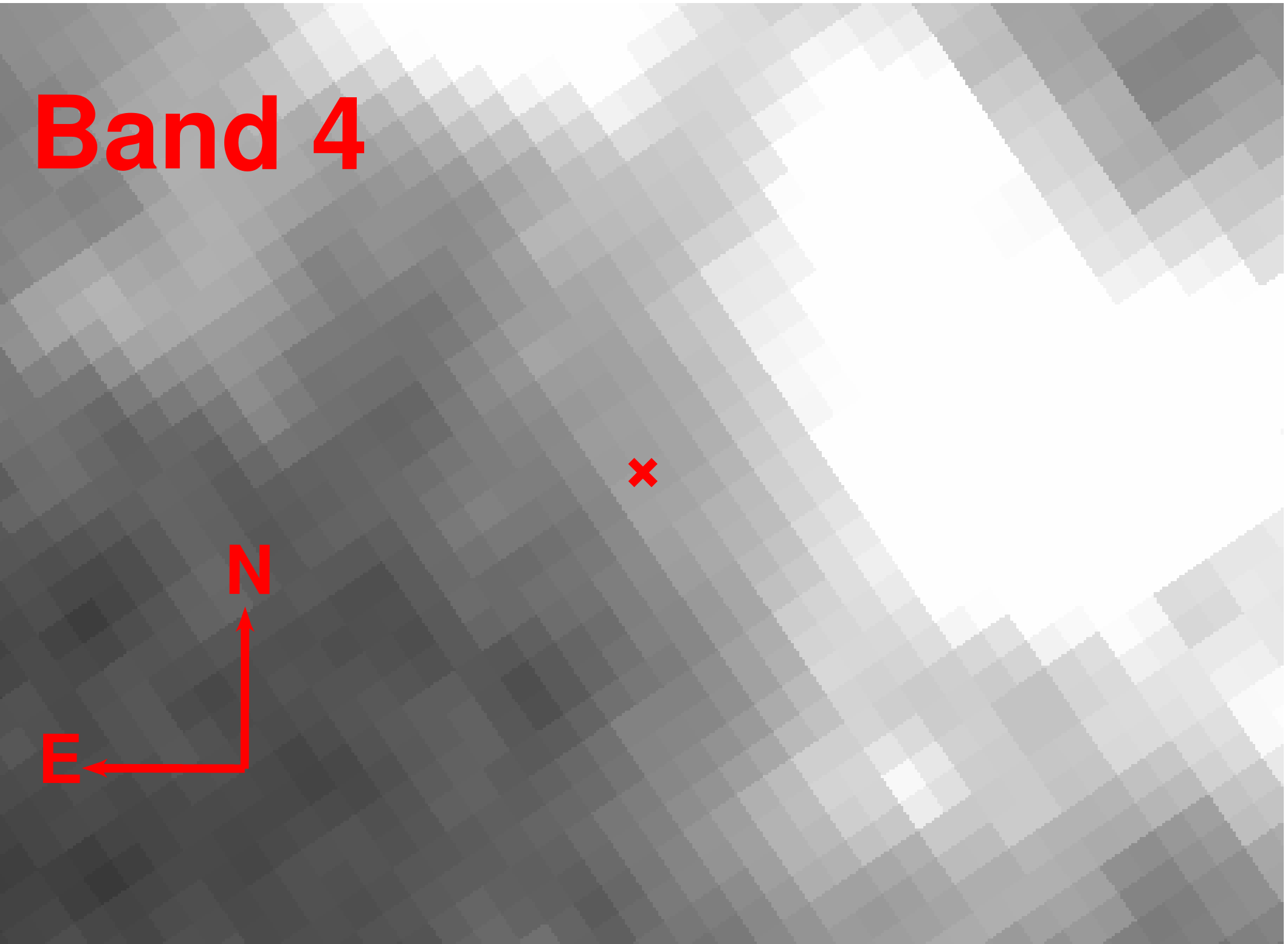}
	\caption{\spitzer/IRAC imaging centred on the Swope $r$-band position of Gaia16cfr.  Bands 1--4 (3.6, 4.5, 5.8, and 8.0 $\mu$m, respectively) are denoted at the top left of each panel.  We detected a point source coincident with the position of Gaia16cfr in Bands 1 and 2 and we have circled these sources with a $2\arcsec.4$ circle in both panels.  We did not detect a point source coincident with Gaia16cfr in Bands 3 and 4.  PSF-fit photometry of the Band 1 and 2 sources and 3$\sigma$ upper limits on the presence of a point source are given in \autoref{tab:spitzer}.}\label{fig:spitzer}
\end{figure}

We also performed relative astrometry between the same Swope $r$-band image and \spitzer/IRAC photometry in order to constrain the position of any pre-outburst counterparts in the IR.  Because \spitzer/IRAC bands trace much cooler, dust-dominated sources, there were typically fewer isolated point sources in each band with which we could anchor the \spitzer\ images; we used 7--12 point sources per band to calculate an astrometric solution.  The astrometric uncertainties in the \spitzer/IRAC WCS solutions were typically $\alpha=0.33$~pixels $\delta=0.32$~pixels, or $\sim 0.2\arcsec$ in both directions.  We show a cutout from each \spitzer\ band centred on the Swope $r$-band position of Gaia16cfr in \autoref{fig:spitzer}.  The ``$x$'' mark shows the Swope position, while the circles in Bands 1 and 2 are centred on point sources extracted using {\tt daophot} and have radii of 2\arcsec.4.  These sources agree with the position of Gaia16cfr to within our astrometric uncertainty.  In Bands 3 and 4, we did not find any point sources within a 2\arcsec.4 radius of the Swope $r$-band position.  Therefore, we calculated 3$\sigma$ upper limits on the presence of any point sources at this position.  The Band 1 and 2 detections, along with the Band 3 and 4 upper limits, are presented in \autoref{tab:spitzer}.

\begin{table}
\begin{center}\begin{minipage}{3.3in}
      \caption{\spitzer/IRAC Photometry}\scriptsize
\begin{tabular}{@{}ccc}\hline\hline
  Wavelength  & Flux Density & Uncertainty \\
  ($\mu$m)    &  ($\mu$Jy)   & ($\mu$Jy)  \\  \hline  
  3.6         &  11.1        & 3.2         \\
  4.5         &  11.7        & 2.7         \\
  5.8         &  $<$29.8     & ---         \\
  8.0         &  $<$10.0     & ---         \\
\hline
\end{tabular}\label{tab:spitzer}
\begin{tablenotes}
      \small 
\item {\bf Note.} Photometry of the Gaia16cfr counterpart obtained on 21 Nov. 2003.
\end{tablenotes}
\end{minipage}
\end{center}
\end{table}

\subsection{Characteristics of the Pre-Outburst Source}

\subsubsection{Optical SED of the Progenitor System}\label{sec:optical}

The pre-outburst source is highly variable, with changes in $F350LP$ as large as $1.4$~mag over 10~days.  This variability, and the possibility that dust absorption at UV/optical wavelengths and emission at IR wavelengths is contributing to the SED, make a precise classification of the underlying source difficult.
 
We first considered a single-component thermal spectrum independently fit to each epoch of the pre-outburst source.  The implied bolometric correction for $F350LP$ is $-0.66$ to $0.02$~mag (with corresponding bolometric magnitudes $-8.9$ to $-11.0$) with a range of temperatures $4100$--$5300$~K for every filter set apart from $F814W$, where we find typical temperatures of $13,000$--$23,000$~K.  This places the Gaia16cfr pre-outburst source either in the range of yellow supergiants such as $\rho$~Cas or firmly in the range of S~Dor-like variables depending on the temperature range we select.  However, the strong wavelength dependence of the single-component SED fitting and the presence of dense CSM as implied by spectra of the outburst event suggest that the pre-outburst SED may contain dust emission or strong line emission, both of which may vary with time.  The brightness at $F658N$ in the 2006 ACS epoch indicates that the $F350LP$ bandpass is contaminated by H$\alpha$ emission, while $F555W$ and $F814W$ have effectively zero throughput near H$\alpha$, so temperature estimates using only two bands are unreliable.  

Next, we considered fitting a single SED to UV/optical emission across multiple epochs.  While the progenitor source is variable and likely has strong contamination from CSM emission, the source has a $F350LP$ ``low'' state over the 79~day period of WFC3 observations near $23.45$~mag (blue dotted line in \autoref{fig:lc}).  Three epochs have $F350LP$ measurements near this low state (i.e., within $\sim0.1$~mag) on JD = 2,457,447.06, 2,457,477.83, and 2,457,488.47.  Therefore, we assume that the overall UV/optical SED of the progenitor source is similar on all three of these epochs and in the three filter pairs from these epochs, which happen to be a single epoch each of $F555W=24.444\pm0.021$ (on JD = 2,457,447.11), $F160W=20.900\pm0.007$ (on JD = 2,457,477.85), and $F814W=23.551\pm0.023$.  Moreover, the $F814W$ measurement is similar to the ACS/WFC $F814W=23.494\pm0.016$ on JD = 2,454,029.39, so we assume that the contemporaneous $F435W$ and $F658N$ measurements are also characteristic of this ``low''-state SED.  We ignore $F160W$ in our initial UV/optical SED fit, as this measurement may be strongly affected by dust in the circumstellar environment of the progenitor source.

We plot the $F435W$, $F555W$, $F350LP$, and $F814W$ UV/optical measurements in \autoref{fig:sed}.  The $F350LP$ and $F814W$ flux densities have been averaged across each measurement in the four epochs we considered, and additional uncertainty ($0.1$~mag and $0.04$~mag, respectively) is added for the standard deviation across all epochs.  Assuming that the $F350LP$ filter contains only emission from the progenitor source and excess H$\alpha$, we subtracted the $F658N$ measurement from $F350LP$ in order to estimate the underlying continuum from the progenitor source.  Accounting for the difference in throughput between ACS/$F658N$ and WFC3/$F350LP$, we find that the subtracted $F350LP$ measurement is $23.9\pm0.1$~mag.

\begin{figure}
	\includegraphics[width=0.5\textwidth]{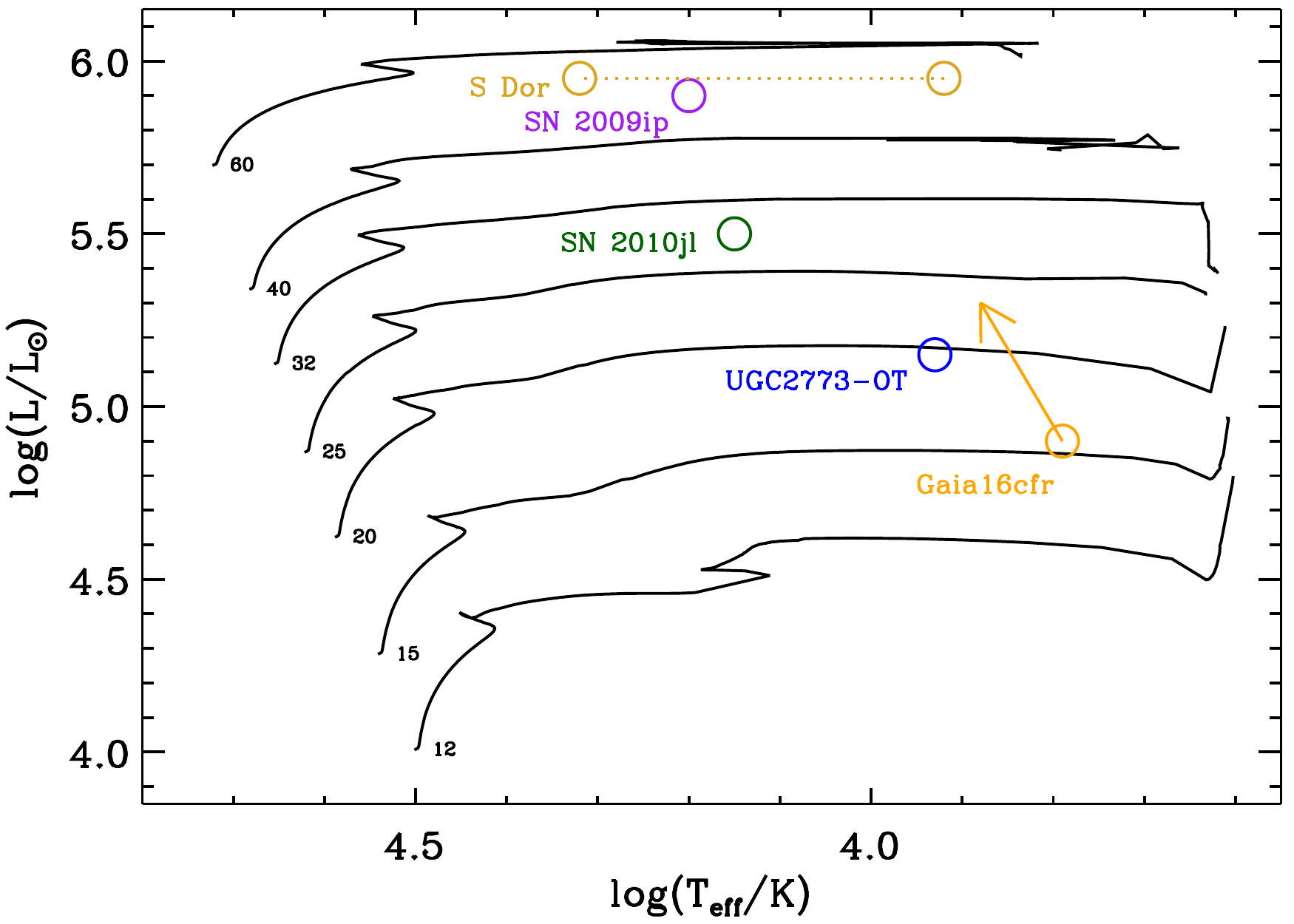}
	\caption{Hertzsprung-Russell diagram showing the derived temperatures and luminosities of SN impostor and SN~IIn progenitor systems.  These include SN~2005gl \citep{gal-yam+07,gal-yam+09,pastorello+17}, SN~2009ip and UGC~2773-OT \citep{smith+10,foley+11}, SN~2010jl \citep{smith+10b,fox+17,dwek+17}, and Gaia16cfr (this paper).  There is no colour information for the progenitor of SN~2005gl, so we adopt the luminosity described by \citet{gal-yam+09} combined with the full range of temperatures for stars at that luminosity predicted by our single-star evolutionary models.  For SN~2010jl, results from \citet{fox+17} suggest that the nature of the progenitor star is only constrained by upper limits from \hst\ and \spitzer, so we instead adopt upper limits for supergiant and LBVs described in Figure 2 of \citet{dwek+17}.  For Gaia16cfr, we display the reddening vector corresponding to $A_{V}=1.0$~mag in a dust shell observed around the progenitor source.  For comparison, we also show S~Dor in both its hot and cool states \citep{lamers95,lamers+98,massey+00,van-genderen+01}, as well as several OPAL single-star evolutionary tracks from \citet{bressan+93} with initial masses indicated.}\label{fig:hr}
\end{figure}

\begin{figure}
	\includegraphics[width=0.5\textwidth]{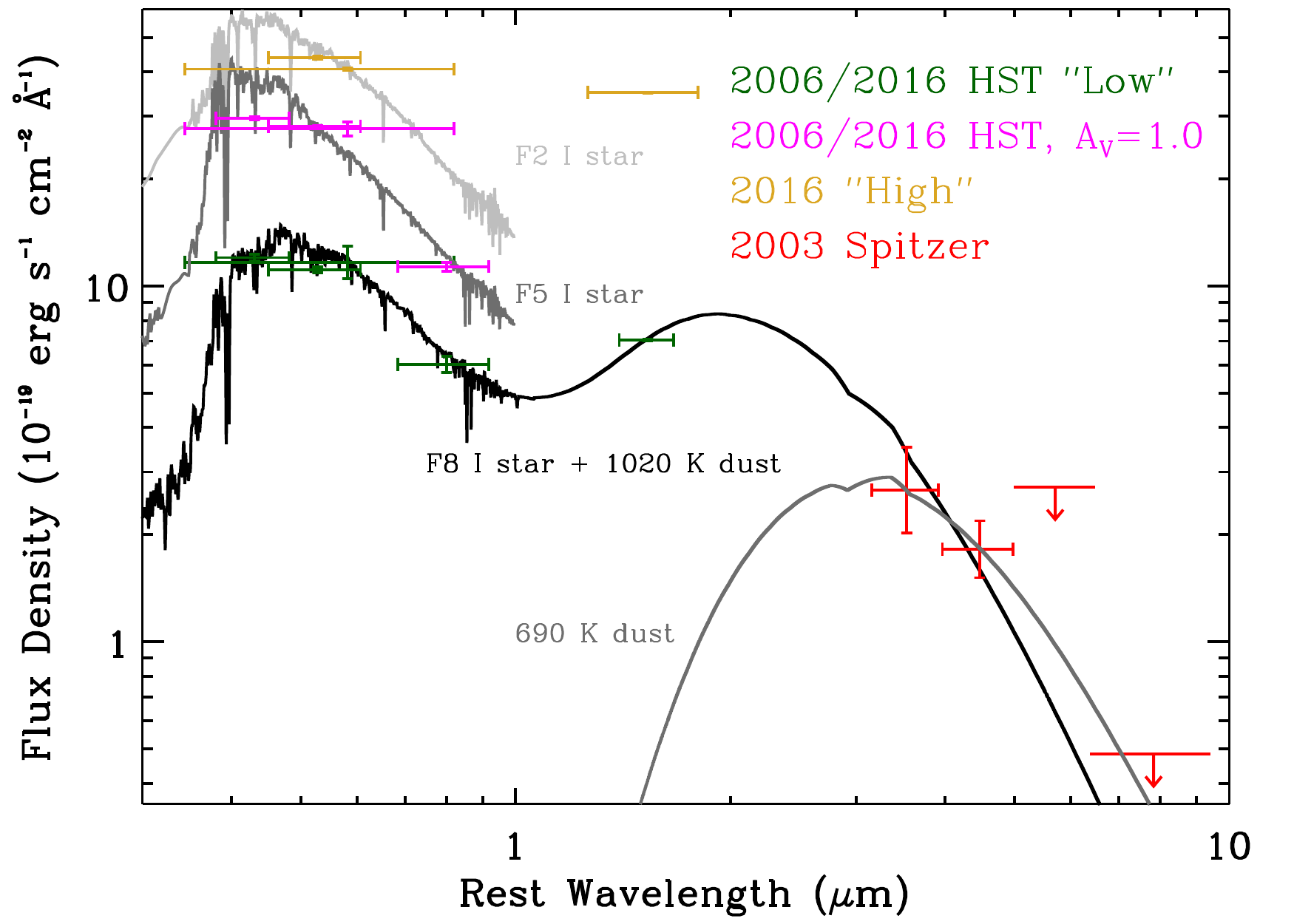}
	\caption{SED demonstrating the \hst\ $F435W$, $F555W$, $F350LP$, $F814W$, and $F160W$ photometry discussed in \autoref{sec:optical} (dark green), as well as the \spitzer/IRAC photometry and upper limits discussed in \autoref{sec:astrom} and presented in \autoref{tab:spitzer} (orange).  The uncertainties in flux density on each point are shown, and the wavelength uncertainties represent the width of the filter from which each photometric point or upper limit was obtained.  We display our overall best-fitting SED in black, which consists of an $18$~M$_{\odot}$ F8~I star combined with a 1020~K dust SED from a $0.1\mu$m graphite grain model, as discussed in \autoref{sec:dust}.  We also present our best-fitting SED for the \spitzer/IRAC photometry alone, which consists of a 690~K dust SED with dust mass absorption coefficients obtained from the same dust model.  Finally, we illustrate an example model of the \hst\ photometry dereddened by $A_{V}=1.0$~mag (cyan) and with the best-fitting stellar SED (an $27$~M$_{\odot}$ F5~I star) overlaid.}\label{fig:sed}
\end{figure}

We fit these UV/optical magnitudes to a range of stellar spectra from \citet{pickles+98}. The best-fitting model is an F8~I star with $\log (T/{\rm K}) = 3.79$ and $\log \left(L/{\rm L}_{\odot}\right)=4.9$, and an initial mass of $18$~M$_{\odot}$ (shown in \autoref{fig:hr}).  The implied photospheric radius from this stellar SED is $260$~R$_{\odot}$ or $1.2$~AU.  This star is much less luminous and cooler than all directly identified SN impostor progenitor stars such as that of UGC2773-OT and SN~2009ip \citep[$\log (L/{\rm L}_{\odot})\geq5.1$ and $\log (L/{\rm L}_{\odot}\approx5.9)$, respectively;][]{smith+10,foley+11}.  However, indirect mass measurements from the stellar population around the SN impostor NGC300-OT \citep[$12$--$25$~M$_{\odot}$;][]{gogarten+09} and upper limits on the luminosity from SN~2008S \citep[$10$--$12$~M$_{\odot}$;][]{prieto+08} indicate that the Gaia16cfr is consistent with the overall SN impostor population.

For comparison, we also plot the pre-outburst SED for the ``high'' state when the magnitudes are near the peak of their variability.  These include $F350LP$ from JD = 2,457,418.83 and JD = 2,457,427.45 (i.e., the second and third epochs of Cepheid data) and the corresponding $F555W$ and $F160W$ magnitudes.  All of these data were significantly brighter during this phase.  Following our analysis in the ``low'' state, we subtract the same $F658N$ magnitude from the $F350LP$ data point for a subtracted measurement of $22.3\pm0.1$~mag.  Although the source SED likely has significant H$\alpha$ emission, which makes an exact spectral classification difficult, we have no reason to believe that the $F658N$ measurement from 2006 is characteristic of the total H$\alpha$ luminosity in the ``high'' state, so this introduces a significant source of uncertainty in our spectral classification of this state.

We plot the ``high''-state flux densities in \autoref{fig:sed} with the best-fitting stellar SED.  We find that the star is significantly more luminous in this state ($\log (L/{\rm L}_{\odot})=5.6$) with a slightly hotter overall SED ($\log (T/{\rm K}) = 3.84$), corresponding to a star with $\sim30$~M$_{\odot}$.  

Significantly, the $F160W$ flux is a factor of $5$ larger (from $10^{4.4}$ to $10^{5.1}~{\rm L}_{\odot}$) than in the ``low'' state, and our stellar SED significantly underpredicts this emission based on the slope from the optical flux densities.  In our model, there is clearly some additional source of IR excess that powers the $F160W$ luminosity.  This trend is extremely unusual, especially as the changes in $F160W$ luminosity occur over a period of $\sim 15$~days.  Whatever source is powering the $F160W$ emission must be compact --- that is, comparable in radius to the underlying optical source.  At the same time, this source must be extremely hot. Even if the source had a characteristic radius of $10$~AU, which implies a large dynamical velocity of $1100$~km~s$^{-1}$ for the 15~day variability, the temperature of an optically-thick IR-emitting source ought to be $\sim2300$~K.  If some of this emission comes from reprocessed light from dust, then for reasonable dust compositions (e.g., graphite/silicate) a large fraction of the dust would be sublimated at these temperatures.  Thus, the variability may be more complicated than changes over the dynamical timescale of a compact circumstellar shell.  We further explore these possibilities, especially the source of the 2003 \spitzer\ emission and the variability in the 2016 \hst\ data, in \autoref{sec:dust} and \autoref{sec:flickering}.

\subsubsection{Infrared Dust Emission and Extinction Toward the Progenitor System}\label{sec:dust}

Assuming the IR SED is dominated by a thermally emitting, spherical dust shell with an optical depth $\tau_{\nu}$ at frequency $\nu$, a blackbody radius $r_{\rm bb}$, a single equilibrium temperature $T$, and a distance $d$, then the dust spectrum follows \citep[as in][]{hildebrand+83}

\begin{equation}
	F_{\nu} = \pi \frac{r_{\rm bb}^{2}}{d^{2}} B_{\nu}(T) \left(1 - \exp(-\tau_{\nu})\right),
\end{equation}

\noindent where $B_{\nu}(T)$ is the Planck function.  In the optically-thin limit, this emission profile follows

\begin{equation}
	F_{\nu}=\pi \frac{r_{\rm bb}^{2}}{d^{2}} B_{\nu}(T) \tau \text{, for }~\tau_{\nu}<<1.
\end{equation}

\noindent Assuming $\tau_{\nu} \approx \rho r_{\rm bb} \kappa_{\nu}$, where $\kappa_{\nu}$ is the dust mass absorption coefficient and $\rho$ is the average density (i.e., $M_{d}/(4/3) \pi r_{\rm bb}^{3}$, with $M_{d}$ being the total dust mass), the optically-thin limit can be expressed as $F_{\nu} \approx M_{d} B_{\nu}(T) \kappa_{\nu}/d^{2}$, as in \citet{fox+10,fox+11}.  IR dust emission around SN~IIn and LBV progenitor stars is usually assumed to be optically thin \citep[as in][]{smith+09,kochanek+11,fox+13}, and we make the same assumption below.

We obtained absorption coefficients for dust grains of a single size and composition from Figure 4 of \citet{fox+10}; however, at IR wavelengths and for dust grains with diameter $<1~\mu$m, the dust-grain size does not affect the overall absorption coefficient. For optically-thin dust composed of $0.1~\mu$m graphite grains, we fit our 2003 \spitzer/IRAC and 2016 \hst/$F160W$ detections to find a total dust mass of $7.7\times10^{-7}$~M$_{\odot}$ with a blackbody temperature of $1020$~K and an overall dust luminosity of $L_{d} = 2.4\times10^{5}~{\rm L}_{\odot}$.  This dust mass is extremely low compared to that observed around virtually all SNe~IIn \citep{fox+11}. Even for a hydrogen-rich mass of CSM with a dust-to-gas mass ratio of $0.01$ \citep[as in][]{fox+10}, the total CSM mass is only about $10^{-4}$~M$_{\odot}$. Furthermore, the blackbody radius implied by the dust luminosity and temperature is $r_{\rm bb} = 72~\text{AU}$.  This final calculation assumes an optically-thick dust shell and is therefore only a lower limit on its size, implying that the 2003 \spitzer\ emission is consistent with a much more extended source than we found in \autoref{sec:optical} \citep[although much more compact than most SNe~IIn with dust shells at 250--4000~AU;][]{fox+11}.

However, as we demonstrate in \autoref{sec:optical-lc}, our assumption that these points form a single, contemporaneous SED may be poor, as the star was highly variable between 2003 and 2016 and the source of the $F160W$ variability in 2016 may be much closer to the progenitor. If we fit only the 2003 \spitzer\ data to a dust SED, we find the best-fitting blackbody temperature is $690$~K with a total dust mass of $4\times10^{-6}$~M$_{\odot}$, comparable to dust masses around SNe~IIn such as SN~2008J \citep{stritzinger+08,fox+11}.  This dust SED implies an overall dust luminosity of $1.3\times10^{5}~{\rm L}_{\odot}$ and a slightly larger blackbody radius of $120$~AU, even larger than we modeled in conjunction with the 2016 data.  While the overall best-fitting parameters are somewhat different in this case, it is clear that the dust shell around the Gaia16cfr progenitor system is relatively low mass and compact compared to dust observed toward most SNe~IIn, although more extended than the 2016 data imply by themselves. However, these properties are in general agreement with post-outburst near-IR spectroscopy of SN~2009ip-12B \citep[with dust mass $4\times10^{-7}$~M$_{\odot}$ and $r_{\rm bb}=120$~AU in][]{smith+13}.

Moreover, even if the dust shell observed in 2003 is unassociated with the emission observed in 2016, the progenitor system may have been episodically ejecting material in the decade before its major outburst and building up its circumstellar envirionment.  We estimate the average mass-loss rate from the Gaia16cfr progenitor system as $\dot{M} \approx 4\pi r_{\rm bb}^{2} \rho_{\rm tot} v$, where $v$ is the wind speed in the CSM ($\sim 250~\text{km~s}^{-1}$, as we discuss in \autoref{sec:halpha}) and $\rho_{\rm tot}$ is the total density of gas and dust (we assume this is $\sim100$ times the dust density).  Given the dust model for the \spitzer\ data, the progneitor system may have been periodically driving $5\times10^{-4}$~M$_{\odot}~\text{yr}^{-1}$ mass loss over a decade before its major outburst.

Finally, although the IR excess is likely associated with some degree of optical extinction, the total amount of extinction is highly uncertain.  For example, if we assume the \spitzer-only model with dust uniformly distributed in a spherical shell, then $\tau_{\nu} = \rho r_{\rm bb} \kappa_{\nu} >> 1$ for typical optical dust mass extinction coefficients $10^{4}$--$10^{5}$~cm$^{2}$~g$^{-1}$.  We have demonstrated that there is a highly variable, H$\alpha$-luminous point source consistent with the position of Gaia16cfr, and so it is unlikely that the source observed in 2006 and 2016 is obscured by this level of dust extinction (e.g., $A_{V}>4$~ mag would require a source with $M_{V}<-11$~mag).

We infer that the dust is either clumpy and unevenly distributed or asymmetric (e.g., in a disk that is at least partly face-on), such that the optical extinction is lower than we might infer from a uniformly distributed dust shell. Therefore, the overall SED of the underlying progenitor source is mostly unconstrained by the IR dust emission and we can only assume that the inferred temperature is a lower limit on the actual source. One way of estimating the total luminosity of the source is to add the total luminosity modeled by the IR dust SED to the optical SED, which implies a total luminosity of $\log(L/{\rm L}_{\odot})\approx5.3$.  Again, this estimate is complicated by the fact that most of the optical SED comes from 2016 \hst\ photometry while the IR SED comes from 2003 \spitzer\ photometry.  However, the $F160W$ photometry suggests that the IR dust luminosity cannot be larger than $\log(L/{\rm L}_{\odot})\approx5.5$ for reasonable dust temperatures \citep[$\sim600$--$1500$~K, as in][]{fox+11}.  If the total luminosity of the progenitor source is $\log(L/{\rm L}_{\odot})\approx5.3$, this would imply $A_{V} = 1.0$~mag with the same $0.1~\mu$m grain graphite dust model, and the most likely stellar SED is an F5~I star with $\log (T/{\rm K}) = 3.88$ and an implied initial mass of $27$~M$_{\odot}$ (see \autoref{fig:sed} and the reddening vector in \autoref{fig:hr}). We also emphasise that in all of our models the implied mass of the progenitor star is low, and so Gaia16cfr is an unlikely candidate for a pulsational pair instability SN.

\subsubsection{Pre-Outburst ``Flickering''}\label{sec:flickering}

\begin{figure}
	\includegraphics[width=0.5\textwidth]{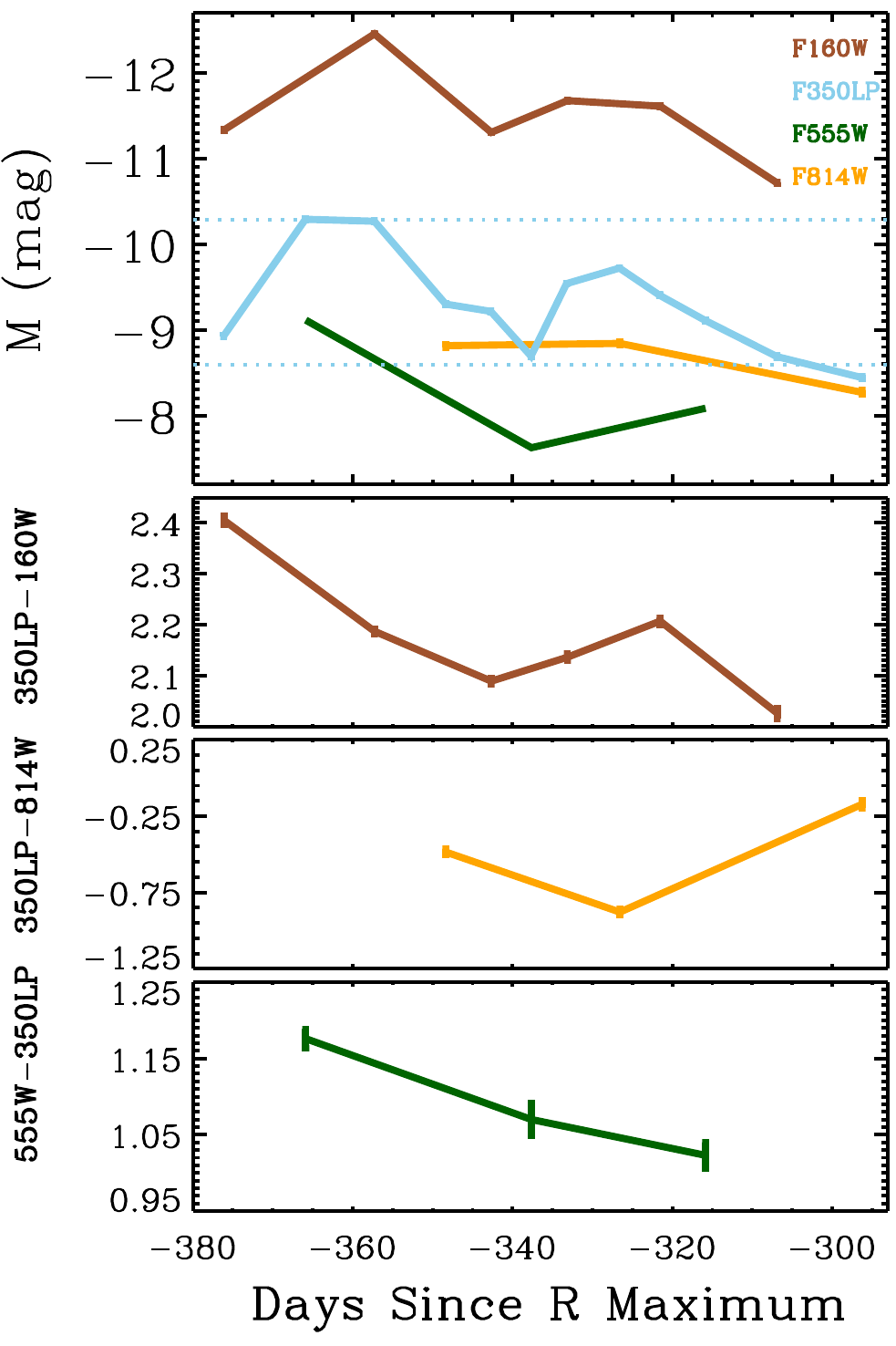}
	\caption{\hst/WFC3 photometry of the pre-outburst counterpart of Gaia16cfr, Jan.--Apr. 2016.  The source exhibited variability on $10$--$30$~day timescales similar to ``flickering'' observed in the pre-outburst light curve of SN~2009ip \citep{smith+10} as well as pre-outburst light curves of other SN impostors \citep[e.g., SN~20154J and SN~2015bh][]{tammann+68,ofek+16,elias-rosa+16,thone+17}.  We indicate dotted lines for the $F350LP$ magnitudes during the ``high'' state and ``low'' state as discussed in \autoref{sec:optical}.  For comparison, we show the $F350LP-F160W$, $F350LP-F814W$, and $F555W-F160W$ colours across our full pre-outburst light curve.  The variation in the optical-IR colour is much smaller and slower than in the overall light curve.  At the same time, the source consistently evolves from redder to bluer colours (especially in $F350LP-F160W$) as it fades.}\label{fig:pre-lc}
\end{figure}

Similar to SN~1954J, SN~2009ip-12B, and SN~2015bh \citep{tammann+68,mauerhan+13,fraser+13,pastorello+13,graham+14,elias-rosa+16}, Gaia16cfr was highly variable within a year of its major outburst (\autoref{fig:pre-lc}).  Over the twelve epochs in which the progenitor source was observed with \hst/WFC3 in $F350LP$, the peak-to-peak variation was roughly $1.8$~mag (a factor of 5.2 in luminosity), with the fastest variations involving $1.4$~mag (factor of 3.6) over 10~days from our first to second epoch.  However, the overall luminosity of the Gaia16cfr progenitor system does not appear to have changed significantly from 2006 to 2016.

We also examined the individual subframes of pre-outburst \hst\ images from 2016, which were usually separated by $8$--$50$~min depending on the filter.  We did not detect any significant variability between the individual images to within the photometric uncertainties (which were typically $30\%$--$50\%$ larger than the combined frames).  This lack of short-timescale variability indicates that the characteristic timescale is much longer than $8$--$50$~min, perhaps as long as the overall $10$--$30$~day timescale observed in the full photometric sequence.  Indeed, the progenitor source fades monotonically in $F350LP$ from JD = 2,457,458.13 to JD = 2,457,488.47, which suggests we are resolving the variability timescale.  

In photometry preceding the outburst of SN~2009ip, \citet{smith+10} referred to this rapid variability as ``flickering'' and referenced similar behaviour in the historical light curve of $\eta$~Car \citep{herschel1847}.  The cause of this variability is perplexing, especially as the timescale appears significantly faster than most processes intrinsic to a progenitor star or its environment. For example, dust extinction plays a role in the overall SED of the Gaia16cfr progenitor system and we noted that there may be some dust destruction in an extended shell. However, \citet{smith+10} remarked that for SN~2009ip, it is unlikely that dust alone could explain such rapid variability for an extended source of emission at $\sim10$~AU, as the dust formation and destruction timescales are longer than the weeks-long timescales observed in the pre-outburst light curve.

In photometry of Gaia16cfr, the apparent photospheric radius from the pre-outburst UV/optical SED of Gaia16cfr is still consistent with that of a typical supergiant ($\sim1$--2~AU) rather than the much larger values required for $\eta$~Car or a progenitor system such as that of SN~2009ip \citep[$\sim10$~AU;][]{davidson+97,smith+10}. It remains plausible that the variability was driven on the dynamical timescale of a progenitor star. Assuming the progenitor were an F8~I star with $18$~M$_{\odot}$, the dynamical timescale is $t_{\rm dyn} \approx {R^{3}/(2 G M)}^{1/2} = 10$~days. Burning instabilities or a wave-driven mechanism \citep[as in, e.g.,][]{fuller+17} could explain the timescale of the observed variability.

Another possibility comes from the wind driven off of the star itself. It has been observed that some LBVs exhibit pseudo-photopheres owing to their optically-thick winds, so the photospheric radius does not reflect the underlying star's hydrostatic radius \citep{groh+08,vink11}.  We have demonstrated that Gaia16cfr has significant IR excess, which is likely from dust emission, and its H$\alpha$ luminosity in 2006 was high ($L_{F658N}=6\times10^{35}~\text{erg s}^{-1}$ corrected for extinction).  If the star is obscured by a significant mass of CSM, then an optically thick, H$\alpha$-emitting wind could explain the timescale of variability.  Assuming a wind velocity of $\sim250~\text{km~s}^{-1}$, then $10$--$30$~day variability implies that the photospheric radius is $\sim1$--$2$~AU, similar to the F8~I model we inferred from the overall pre-outburst photometry.  This value is also consistent with the observed photospheric radius of LBVs such as S~Dor \citep{lamers95,lamers+98,van-genderen+01}, as well as models of S~Dor-like LBVs during their ``maximum'' phase \citep[i.e., the outbursting phase;][]{leitherer+89}.

In \autoref{fig:pre-lc}, we also plot the colours of the Gaia16cfr pre-outburst source during this ``flickering.'' There is some variation in the IR excess $F350LP-F160W$ ($0.38$~mag peak-to-peak) over the period of our observations, although it is much slower and weaker than the overall variation in both the optical and IR bands.  The source is simultaneously becoming brighter in optical bands ($F350LP$, with $>5\%$ throughput from $3327$--$9631$~\AA) and in $F160W$.  In general, the source appears reddest when it is close to its maximum around day $-380$ to day $-360$, which is generally consistent with S~Dor-like variability.  However, the trend in $F350LP$ and $F160W$ indicates that the actual luminosity of the optical/IR source is changing rather shifting from an IR-dominated SED to one that is more optically bright.  Interaction between a strong, optically-thick wind and a compact, dusty shell of CSM is an obvious candidate for this additional luminosity, and it agrees with the overall timescale of variability as we discussed above.  Again, dust destruction is likely to occur at this phase, especially if the UV/X-ray emission in the system was enhanced from circumstellar interaction.  The overall trend toward bluer optical-IR colours suggests that the source of IR emission may have been getting hotter or less massive (or some combination of the two) while the optical SED was enhanced by strong continuum emission and Balmer lines from circumstellar interaction.

This interpretation suggests that the star was periodically driving precursor outbursts before the major outburst in December 2016.  Moreover, it is curious that SN~1954J, SN~2009ip-12B, and SN~2015bh all exhibited significant variability on weeks-long timescales roughly a year before their major outbursts \citep{tammann+68,smith+10,thone+17}.  Any physical mechanism that can account for the major outburst must also explain why the star undergoes these precursor events and why they are timed to within years or months of the outburst itself.

\begin{figure*}
	\includegraphics[width=\textwidth]{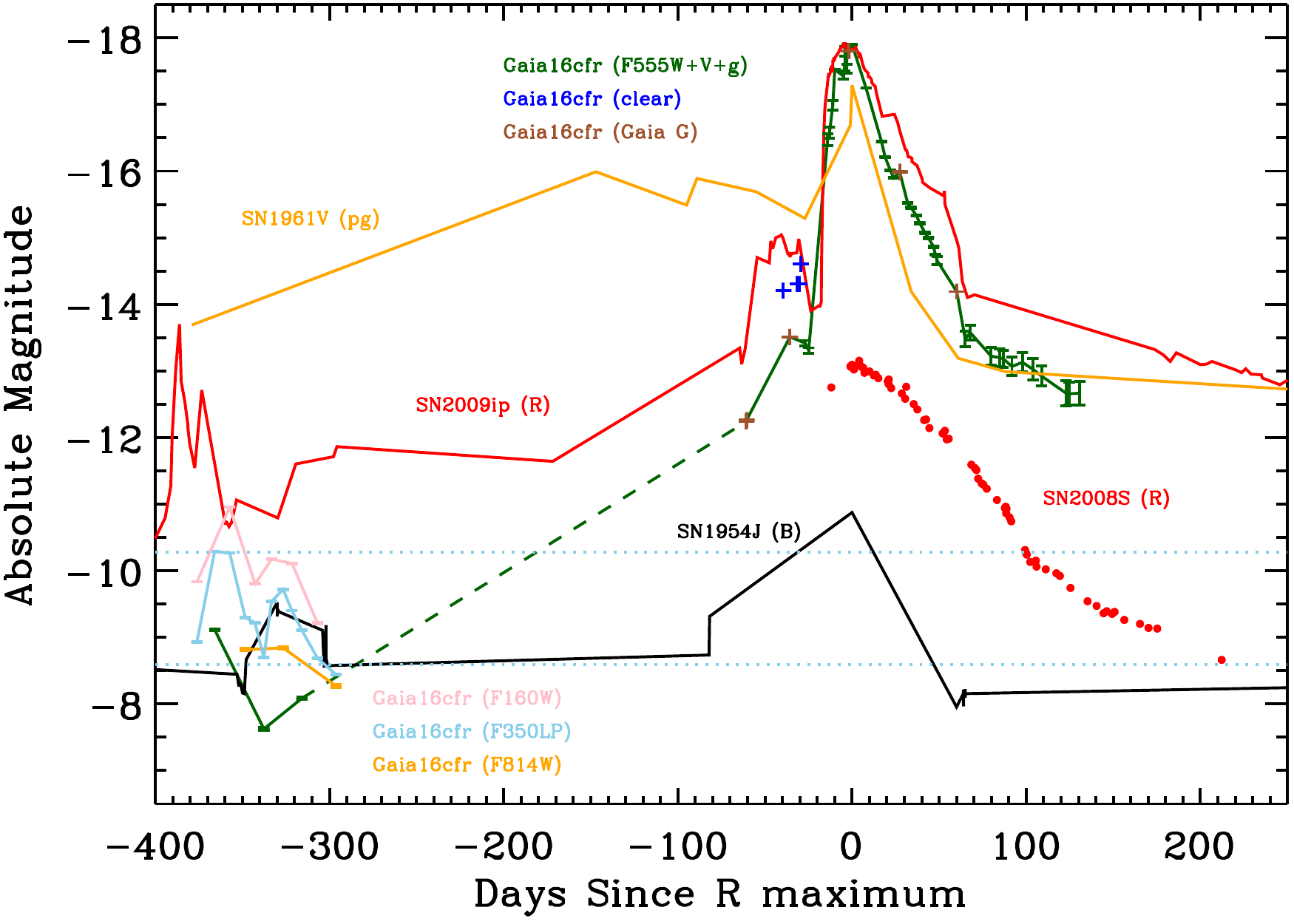}
	\caption{The optical light curve of Gaia16cfr from pre-outburst through our full $g$-band Swope and $V$-band EFOSC2 light curve, and including Gaia photometry and reported ``clear'' photometry \citep{atel9937,atel9938}.  For comparison, we overplot the $B$-band light curve of SN~1954J \citep{tammann+68}, the ``photographic'' (pg) light curve of SN~1961V \citep{zwicky+64}, the $R$-band light curve of SN~2009ip-12B \citep{fraser+13,mauerhan+13,pastorello+13,graham+14}, and the $R$ light curve of SN~2008S \citep{prieto+08,smith+09}.  Around 380 days before $R$ maximum, SN~2009ip exhibited significant variability \citep[the ``2011 eruptions'' and the SN~2009ip-12A event immediately before the rapid rise to peak;][]{pastorello+13,graham+14} interpreted as precursor outbursts before the SN~2009ip-12B event.  Gaia16cfr also exhibited significant variability on roughly the same timescale before outburst.  It reached the same peak magnitude as SN~2009ip-12B and declined at roughly the same rate.  We highlight precursor variability in Gaia16cfr, including the difference between the ``low'' state and ``high'' state of the pre-outburst source in $F350LP$ (dotted blue lines), which is separated by $1.7$~mag.}\label{fig:lc}
\end{figure*}

\subsection{Optical Light Curve of the Outburst}\label{sec:optical-lc}

In \autoref{fig:lc}, we compare the absolute magnitude of Gaia16cfr in the $V$, $F555W$, ``clear,'' and Gaia $G$ bands to photometry from several other objects. These data include $R$-band photometry of the LBV outburst SN~2008S\footnote{These data and spectra in \autoref{sec:spectra} come from \url{sne.space}.  See also \citet{guillochon+16}.} \citep{smith+09}, $R$-band photometry of SN~2009ip-12B \citep{fraser+13,pastorello+13,graham+14}, $B$-band photometry of SN~1954J \citep{tammann+68}, and unfiltered photographic (``pg'') photometry of SN~1961V \citep{zwicky+64}.  These light curves are all corrected for distances and extinction using values given in each reference.

Before maximum light, Gaia16cfr became significantly brighter than the pre-outburst photometry. Within the 35~days before maximum light, the {\it Gaia} $G$-band photometry brightened by $4.5$~mag, and there appears to have been a gradual increase in luminosity roughly $30$~days before maximum light as reported by \citet{atel9937}. As our EFOSC $V$-band photometry demonstrates, the source was declining in magnitude within $25$~days of optical maximum and roughly at the same $V$-band luminosity and timescale as SN~2009ip-12B.  Immediately after this decline and within the span of $11$~days from 6 Jan. to 17 Jan. 2017, Gaia16cfr increased in luminosity by $3$~mag and continued to rise to its peak magnitude around 31 Jan. 2017 (\autoref{fig:zoomed}). These data suggest that Gaia16cfr was discovered when it was undergoing a precursor outburst, similar to the pre-maximum variability observed from SN~2009ip-12B \citep[i.e., the SN~2009ip-12A event in][]{pastorello+13,graham+14}.  Although this rise is not as tightly constrained as the SN~2009ip-12B event, the similarities between these objects strongly imply that their light curves followed a comparable rise.

\begin{figure}
	\includegraphics[width=0.5\textwidth]{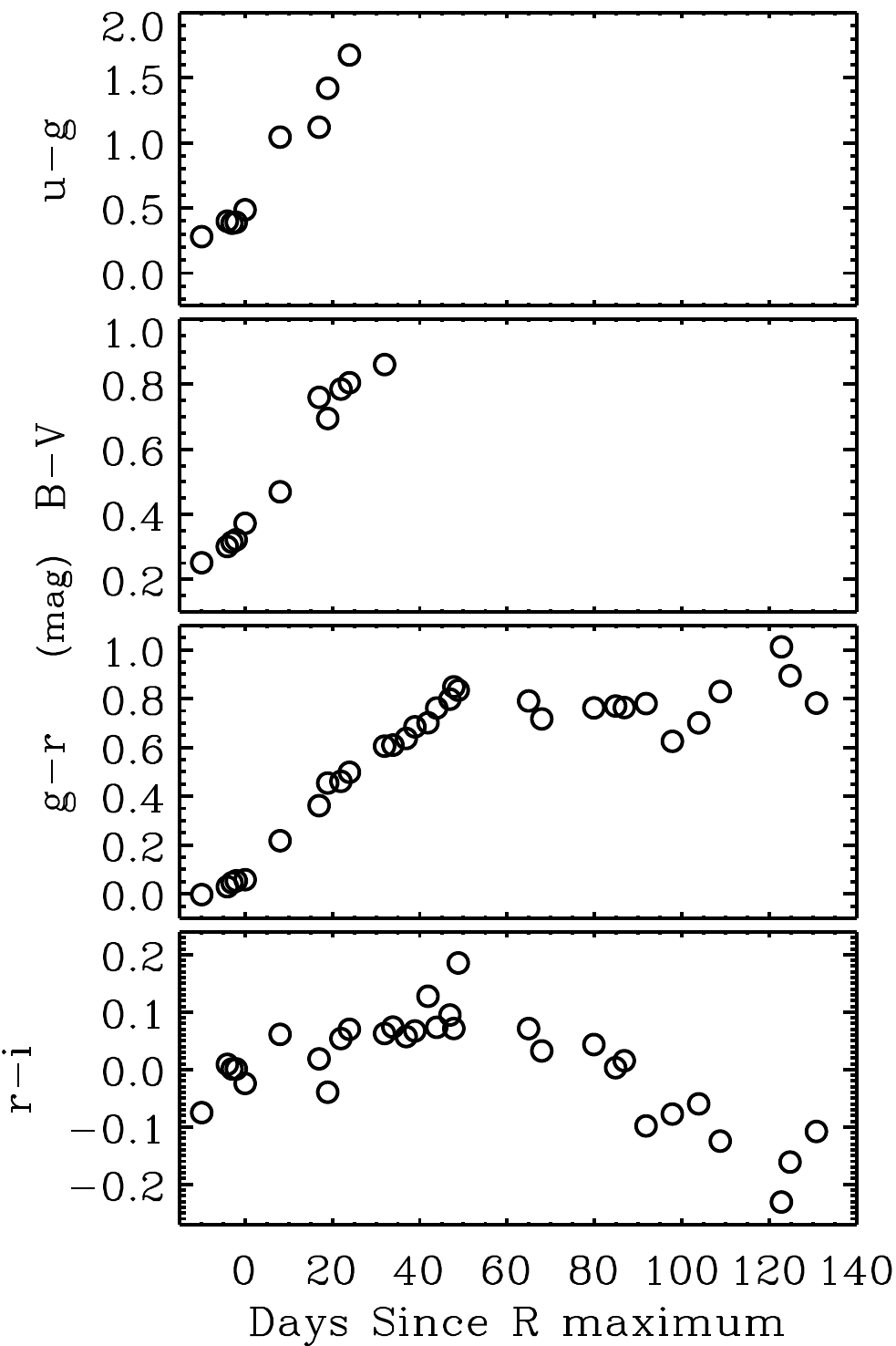}
	\caption{The $u-g$, $B-V$, $g-r$, and $r-i$ colour curves of Gaia16cfr.  The transient steadily evolved to redder colours through optical maximum brightness, with the largest changes occurring in the $u$ band.  At late times, the source became slightly bluer as seen in $g-r$, perhaps due to enhanced continuum emission from further interaction with CSM.  This timescale agrees with the enhancement in the blueshifted H$\alpha$ profile that we discuss in \autoref{sec:halpha}.}\label{fig:colors}
\end{figure}

The similarity between Gaia16cfr and the SN~2009ip-12B is most apparent near peak luminosity (\autoref{fig:zoomed}). Both of these events exhibited $r$-band peaks of $-18$~mag.  Gaia16cfr became steadily redder over time (\autoref{fig:colors}), with the largest changes occurring in the $u$ band throughout our light curve.

\begin{figure}
	\includegraphics[width=0.5\textwidth]{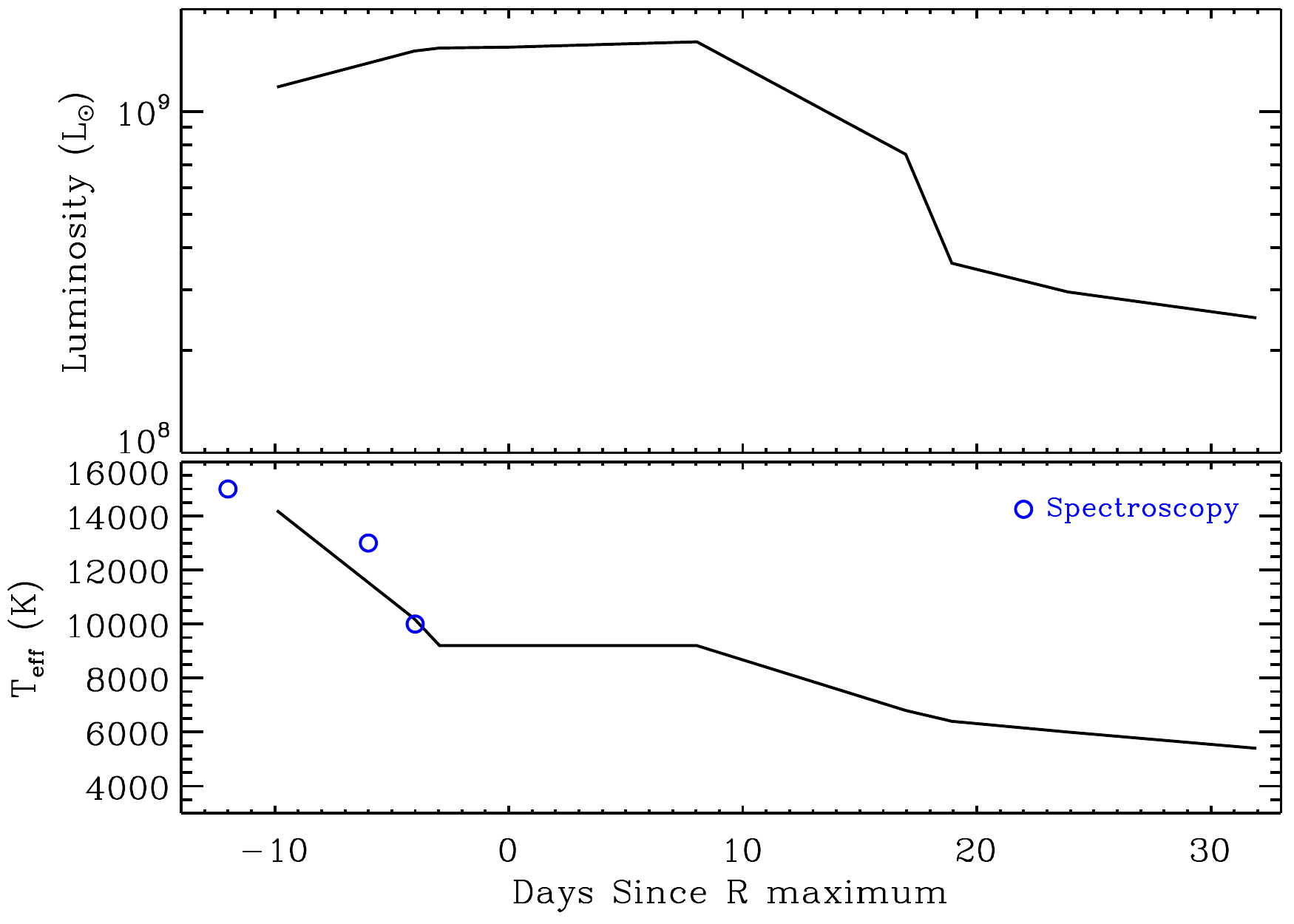}
	\caption{The derived luminosity and temperature of Gaia16cfr from our Swope optical light curve.  The transient peaks at $\sim 1.6\times10^{9}~{\rm L}_{\odot}$, with a slow decline and a clear plateau in optical light starting $20$~days after optical maximum.  We also plot the temperature derived from our best-fitting blackbody to the $uBVgri$ photometry.  For comparison, we show the best-fitting temperature from our earliest three epochs of spectroscopy as blue circles (see \autoref{sec:continuum}).}\label{fig:lumtemp}
\end{figure}

\begin{figure}
	\includegraphics[width=0.5\textwidth]{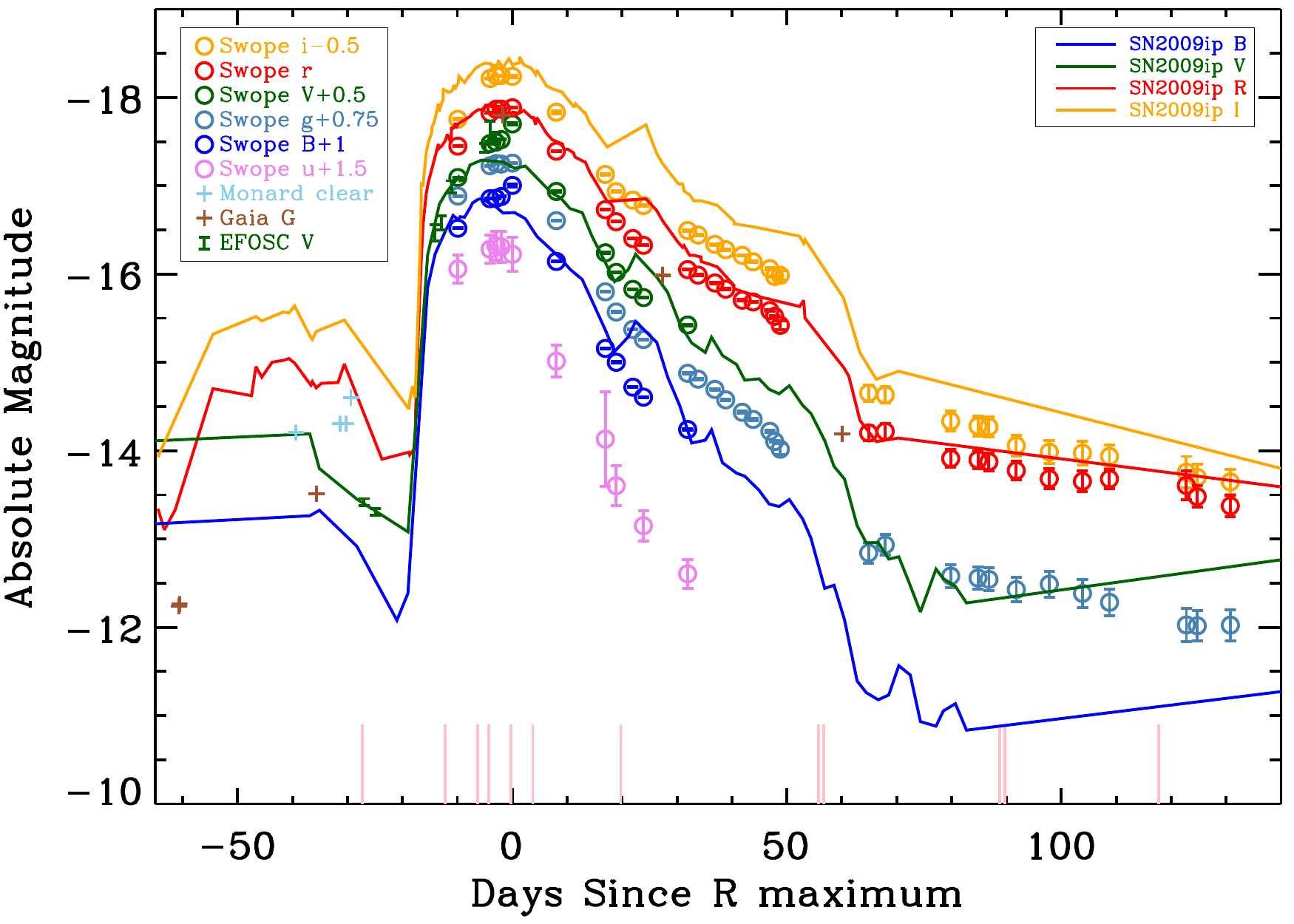}
	\caption{Full Swope+EFOSC2+{\it Gaia}+``clear'' light curve of Gaia16cfr.  We show photometry of SN~2009ip-12B for comparison \citep{fraser+13,mauerhan+13}.  Gaia16cfr and SN~2009ip-12B reached roughly the same absolute magnitude in $BVRI$ and declined at the same rate.  Both events exhibited a plateau in their light curves in all bands 60~days after $r$-band maximum.  This plateau may have begun earlier, as shown by our derived bolometric luminosity in \autoref{fig:lumtemp}.}\label{fig:zoomed}
\end{figure}

To estimate the bolometric luminosity of Gaia16cfr, we fit a blackbody spectrum to the Swope $uBVgi$ photometry, excluding the $r$ band photometry as Balmer emission may bias our fits. We used this blackbody spectrum to derive a temperature and $r$-band bolometric correction and applied this value to our $r$-band magnitude for each epoch of Swope photometry. In this way, we simultaneously fit the thermal component represented in $uBVgi$ bands and Balmer component, which is mostly contained in the $r$ band. Our earliest photometry corresponds to 10~days before $r$-band maximum (day $-10$; we indicate the phase of our light curve and spectroscopy with day number relative to maximum light in the $r$ band) and the best-fitting temperature at this time was $14,000$~K, with an implied luminosity of $1.1\times10^{9}~{\rm L}_{\odot}$ and a photospheric radius of $\sim26$~AU.  Gaia16cfr peaks at $\sim 1.6\times10^{9}~{\rm L}_{\odot}$ ($M_{\rm bol}=-18.3$~mag) and is still about $9200$~K (\autoref{fig:lumtemp}) at this time (which roughly agrees with our spectra in \autoref{sec:continuum}).

The photospheric radius we measure at optical maximum is about $70$~AU, which is in agreement with our estimates of the dust shell at $70$--$120$~AU in earlier epochs.  We infer that the bulk of the optical luminosity comes from an interaction between the ejecta and dusty shells of CSM ejected by the star in precursor outbursts.  This interpretation is supported by the evolution of Gaia16cfr before and near optical maximum, which indicate that the luminosity of the source rose sharply within $\sim14$--$25$~days of maximum, most likely when the high-velocity outburst material encountered the inner radius of a circumstellar shell.  Indeed, the rise in {\it Gaia} $G$-band emission in Dec. 2016 and subsequent decline in EFOSC $V$-band emission in Jan. 2017 was likely the stellar outburst itself, and the interaction-powered light curve only began once the outburst ejecta caught up to the CSM.

At what velocity was the bulk of these ejecta traveling?  If we track the radius of the photosphere over the $\sim10$~days from our first photometry point to optical maximum, we find $v\approx 7500~\text{km~s}^{-1}$, although this value is uncertain and likely larger than the ejecta velocity as the photosphere traces the forward shock.  In SNe~IIn, the forward-shock velocity as inferred from the evolution of optical, radio, and X-ray emission is typically $2$--$4$ times the ejecta velocity \citep{pooley+01,chandra+15,chevalier+16,smith+17}. We conclude that most of the ejecta from Gaia16cfr were moving significantly slower than $7500~\text{km~s}^{-1}$, which may indicate that it is slower than most core-collapse SNe \citep[e.g.,][]{zampieri+03,hamuy+03,valenti+09}.

Integrating the inferred bolometric emission over the full range of dates (day $-10$ to day 31) for which we have $u$-band measurements suggests that Gaia16cfr radiated a total energy of $\sim10^{49}~\text{erg}$.  This total radiated luminosity is comparable to that of many SNe~IIn \citep[e.g., PTF12cxj, SN~2010mc, SN~2011ht;][]{ofek+13,mauerhan+13b,smith+14,ofek+14a,ofek+14b}.  For a relatively low efficiency of converting kinetic energy to optical luminosity ($E_{\rm rad}/E_{\rm k}<0.1$), Gaia16cfr could be consistent with a low-energy core-collapse SN with $E_{\rm k}\approx 10^{50}$~erg.  However, it is unclear whether this low efficiency holds true. For SN~2009ip-12B, spectropolarimetry indicated that the outburst was evolving into an aspherical circumstellar environment, likely arranged in a ring \citep{mauerhan+14}.  If the circumstellar environment of Gaia16cfr were arranged in such a way, a small fraction of the ejecta might be encountering circumstellar material and the energy in the ejecta could be very high.  But from the optical light curve alone, we can only interpret the total integrated luminosity as a lower limit on the explosion energy.

After maximum light, the evolution of the Gaia16cfr light curve is nearly identical to that of SN~2009ip-12B, with $\sim 0.05$~mag day$^{-1}$ decline rate after peak in optical bands followed by a period of more rapid decline and a plateau after day 60.  This plateau may have begun even earlier, as shown by our derived bolometric luminosity (\autoref{fig:lumtemp}), but the steadily cooling photosphere continued to shift emission to redder bands, causing an apparent decline in optical light.  As the plateau begins, the $g-r$ colour levels off (\autoref{fig:colors}). This same behaviour was observed from SN~2009ip-12B at later times when the UV/optical light curve flattened and gradually rebrightened, occurring first in redder bands \citep{fraser+13}.

It was hypothesised that the timescale of this rebrightening after optical maximum in SN~2009ip-12B was consistent with an interaction between material moving at $\sim500~\text{km~s}^{-1}$ and ejecta from the 2012 eruption moving at $4500~\text{km~s}^{-1}$ \citep{graham+14}.  Although we have demonstrated that CSM was present around the Gaia16cfr progenitor system in 2016, we do not have a constraint on when this material was ejected.  Even assuming the slowest CSM velocities for material around SN impostors or SNe~IIn \citep[e.g., 75--200~km s$^{-1}$ as in NGC300-OT and SN~2005ip;][]{bond+09,smith+09}, the longest timescale for a dust shell at $70$--$120$~AU is only $\sim8$~yr.  As we demonstrate in \autoref{sec:halpha}, the narrow, Lorentzian H$\alpha$ profile in our spectra is consistent with a CSM FWHM velocity of $250$~km s$^{-1}$. Therefore, the compact dust shell at 70--120~AU was likely ejected within $1$--$2$~yr of the outburst.

It is also possible that this plateau is intrinsic to the explosion.  \citet{lovegrove+17} predict that for very low-energy SNe of stars in the $15$--$25$~M$_{\odot}$ mass range, the outer hydrogen envelope will become unbound and produce a plateau with a duration that scales roughly as $t \sim E^{-1/6}$ and a luminosity that scales as $E^{5/6}$.  Perhaps as the interaction region becomes optically thin, we are seeing through to the outer hydrogen envelope (or some fraction of the envelope ejected by the progenitor star) that is producing radiation mostly through recombination.  The recombination luminosity will be relatively high ($10^{41}$--$10^{42}$~erg~s$^{-1}$ with $M_{\rm bol}=-13.8$--$-16.2$~mag) for explosions with $E_{k}=10^{49}$--$10^{50}$~erg.  This range roughly agrees with the behaviour of Gaia16cfr at very late times where our Swope $ri$ magnitudes are of order $-14$~mag, although the light curve is steadily becoming fainter in $gri$.  Late-time photometry of Gaia16cfr will be critical to determine the timescale and overall luminosity of this plateau in order to investigate its underlying mechanism.

Our pre-outburst $F160W$ observations provide a constraint on the presence of dust in the environment of Gaia16cfr, but our optical light curve and spectra imply that any dust in the 2003 \spitzer\ data is {\it unassociated} with the configuration of the system in the pre-outburst 2016 observations.  Although the 2003 \spitzer\ data are still consistent with a relatively compact dust configuration as we demonstrated in \autoref{sec:dust}, it is likely that this material was ejected in a previous outburst via some mechanism that periodically ejected shells of CSM.  Thus, the Gaia16cfr pre-outburst and post-outburst data indicate that the progenitor system underwent multiple recent ejections before its major outburst in Dec. 2016.

\subsection{Spectroscopic Morphology of the Outburst}\label{sec:spectra}

\begin{figure}
	\includegraphics[width=0.5\textwidth]{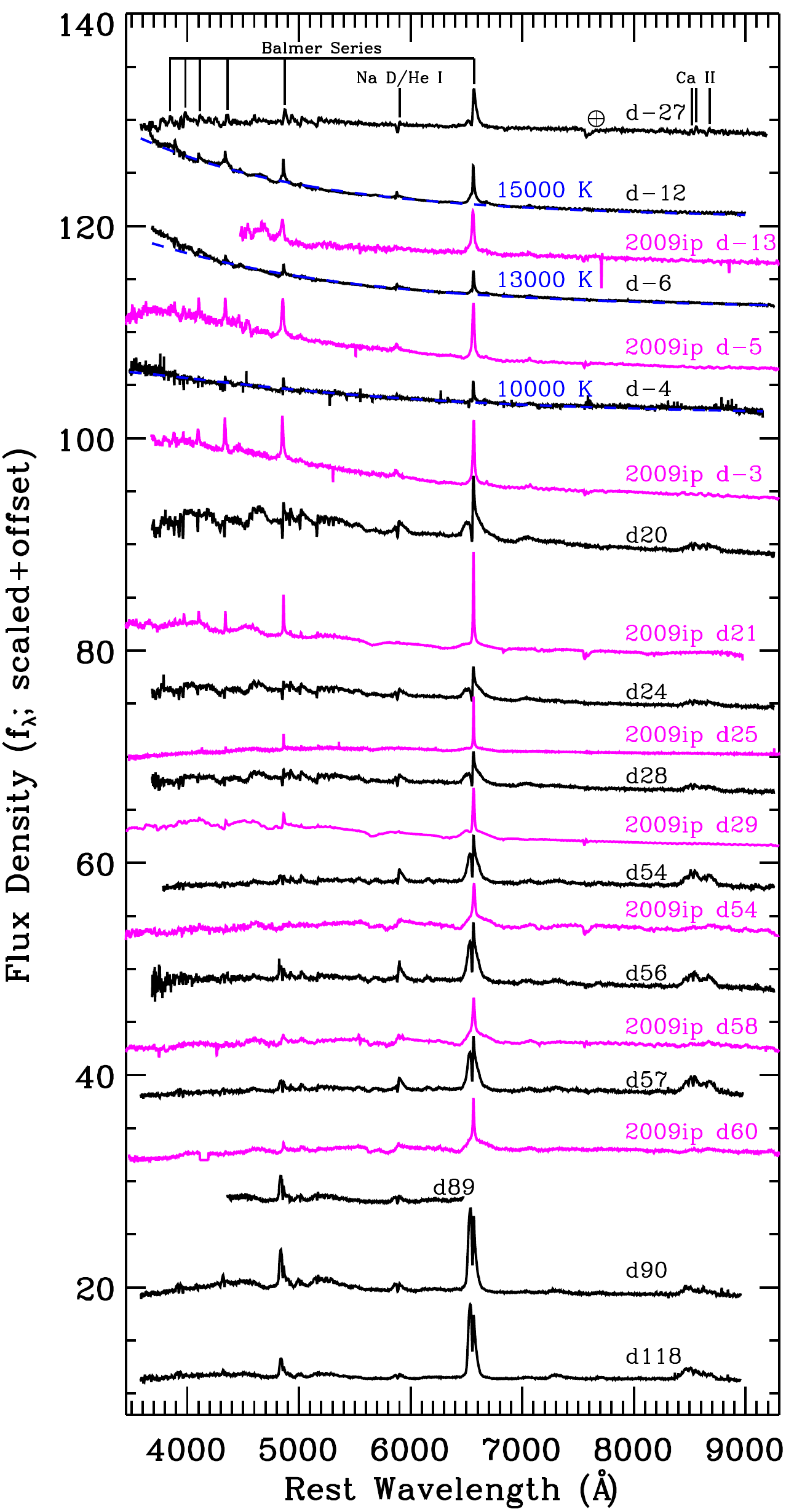}
	\caption{Our full spectral series of Gaia16cfr.  We label each spectrum d\#\# according to its date relative to $r$-band maximum.  Several spectroscopic features, including the Balmer series, Na~D absorption, and He\I\ emission, are noted.  We also mark a telluric feature in the first spectral epoch.  Our day $-12$ spectral epoch exhibits strong blue continuum emission that is well-fit by thermal continuum emission with $T=15,000$~K.  For comparison, we overplot several spectra of SN~2009ip-12B from \citet{mauerhan+13,margutti+14,pastorello+13,graham+14,childress+16}. Note that a few of the SN~2009ip spectra, such as that obtained on day 25, exhibit a deficit of flux at short wavelengths, probably because they were not obtained at the parallactic angle \citep{filippenko82}.}\label{fig:spectra}
\end{figure}

We show our full spectroscopic series in \autoref{fig:spectra}.  These spectra span a wide range of timescales in the outburst from well before (day $-27$) to after maximum light (day $118$).  It is curious that, although Gaia16cfr is photometrically very similar to SN~2009ip-12B as discussed in \autoref{sec:optical-lc}, there are many spectroscopic differences between these two events, especially in the overall profile of H$\alpha$.  Below, we highlight several significant features and associate them with the morphology of Gaia16cfr at various epochs, specifically the evolution of its ejecta and their interaction with the circumstellar environment around Gaia16cfr.

\subsubsection{Thermal Continuum Emission}\label{sec:continuum}

The characteristic blue continuum emission associated with CSM interaction is most obvious in our day $-12$ epoch, where we find it is best fit by a blackbody spectrum with $T=15,000~\text{K}$, cooling to $13,000$~K at day $-6$ and $10,000$ at day $-4$.  This temperature is poorly constrained because the peak of the emission is well into the UV, and the continuum in the blue/UV part of our day $-12$ spectrum is dominated by Fe absorption.  However, we can infer that the overall temperature is high.  This is curious, as there is no clear signature of high-ionisation species such as He\II\ $\lambda\lambda$4686, 5412, C\IV\ $\lambda$5801, or N\IV\ $\lambda\lambda$5047, 7123.  These high-ionisation lines are often observed in ``flash spectroscopy'' of SNe soon after explosion \citep{khazov+15}, but are entirely absent in our spectra.

As we demonstrate in \autoref{fig:lumtemp}, the temperature of the Gaia16cfr photosphere was already cooling starting from the day $-12$ epoch.  It is possible that the day $-12$ spectrum was observed at a special time in the evolution of Gaia16cfr. That is, the shock interaction between ejecta and CSM had not cooled significantly, but given high electron densities in the shocked region, the recombination timescales for the highest ionisation species were short and corresponding line emission was not present. This evolution was observed in the Type IIb SN~2013cu, where the cooling envelope phase after shock breakout was accompanied by high-ionisation species as seen in spectra roughly half a day after core collapse \citep{gal-yam+14}.  However, within 6~days of core collapse, the SN~2013cu spectrum evolved into a relatively featureless, but still extremely blue, continuum.  Even if high-ionisation species were present at a relatively low level in Gaia16cfr, strong continuum emission might decrease the S/N of a detection, as has been noted in SN~IIn and SN~Ia/IIn spectra \citep{smith+07,fox+15,kilpatrick+16}.

Many SN impostors exhibit strong thermal continuum emission and Lorentzian H$\alpha$ profiles in optical spectra \citep[e.g., SN~2008S, SN~2009ip, UGC2773-OT, SN~2015bh;][]{smith+09,smith+10,mauerhan+13,elias-rosa+16}, and the temperature observed for Gaia16cfr near peak is comparable to the bluest examples (e.g., SN~2015bh was roughly $20,000$~K near peak). Moreover, this temperature exceeds the threshold of the value at which dust grains can survive. Given our prediction of a relatively compact dust shell surrounding the progenitor system and the timescale on which ejecta traveling at $1500$--$2000~\text{km~s}^{-1}$ could encounter this dust, it is likely that a significant fraction of the pre-outburst dust was vaporised during this phase, allowing us to peer through some of the CSM to the inner ejecta regions (e.g., where [Ca\II] is formed). The low continuum levels observed in the blue after our second epoch (as seen in \autoref{fig:spectra}), as well as the sharp dropoff in the $u$-band luminosity after peak brightness (\autoref{fig:lc}), indicate that the opacity in the outflowing material produced by electron scattering likely decreased significantly after this phase. The drop in electron scattering and dust absorption suggests that the spectroscopic morphology of the later epochs is dominated by features originating deeper inside of the outburst.

\subsubsection{Calcium Emission and Absorption}

\begin{figure}
	\includegraphics[width=0.47\textwidth]{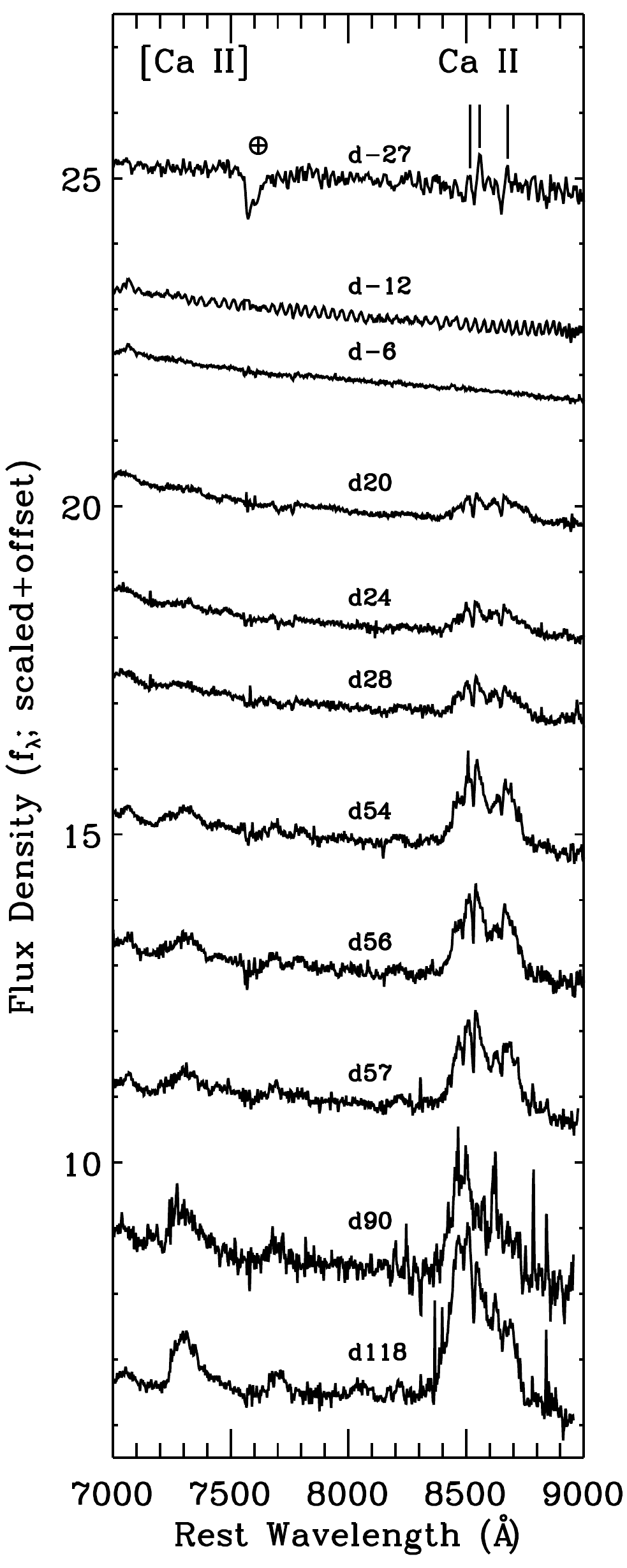}
	\caption{A cutout of our full spectral series focused on the [Ca\II] $\lambda\lambda$7291, 7323  and Ca\II\ IR triplet lines between $7000$ and $9000$~\AA.  The Ca\II\ features are present, but relatively weak and narrow in the first spectral epoch.  These lines become significantly stronger and broader in the two post-maximum epochs.  On the other hand, [Ca\II] emission is relatively weak \citep[especially compared to that of SN impostors such as SN~2008S and UGC2773-OT;][]{prieto+08,smith+09,smith+10}.  We only detect significant [Ca\II] starting after day $54$ (54 days after optical maximum).}\label{fig:caii}
\end{figure}

Significant Ca\II\ IR triplet emission is apparent in the day $-27$ epoch as well as times beyond day $20$ (\autoref{fig:caii}).  We do not see any significant [Ca\II] $\lambda\lambda$7291, 7323 emission until much later epochs. This combination of strong Ca\II\ IR triplet emission with little or no [Ca\II] emission is in stark contrast to many SN impostors, notably SN~2008S, NGC300-OT, and UGC2773-OT \citep{smith+09,berger+09,bond+09,smith+10}.  \citet{berger+09} found relatively narrow [Ca\II] emission in NGC300-OT and noted that this line likely requires a physically distinct and lower density region with a high electron fraction to excite the forbidden emission.

We infer that the [Ca\II] emission traces ejecta unshocked by the CSM in the interaction region.  However, the Ca\II\ IR triplet forms in the high-density CSM itself, and as the evolution from our day $-12$ to late-time emission demonstrate (\autoref{fig:caii}), the Ca\II\ feature becomes significantly broader in the post-maximum phase.  This evolution is likely caused by shock acceleration of the CSM by the outburst.  We are unable to see into the low-density, unshocked ejecta where [Ca\II] forms until the latest spectral epochs.

The fact that the [Ca\II] emission is generally weaker than the Ca\II\ IR triplet emission is perhaps consistent with our interpretation of the pre-outburst dust SED and a relatively compact, dense dust shell.  Unlike other SN impostors where the initial dust shell is relatively extended or diffuse \citep[e.g., SN~2008S, where the dust shell was predicted to have $L\approx 8\times10^{4}$~L$_{\odot}$ with a radius of 230~AU;][]{prieto+08}, the initial compact configuration of Gaia16cfr led to a scenario in which most of the ejecta behind the interaction region is either in a dense, shocked region or still obscured.  As the transient evolves, we expect the ratio of [Ca\II] to Ca\II\ IR triplet emission to continue to increase.

\subsubsection{H$\alpha$ Profile}\label{sec:halpha}

\begin{figure}
	\includegraphics[width=0.49\textwidth]{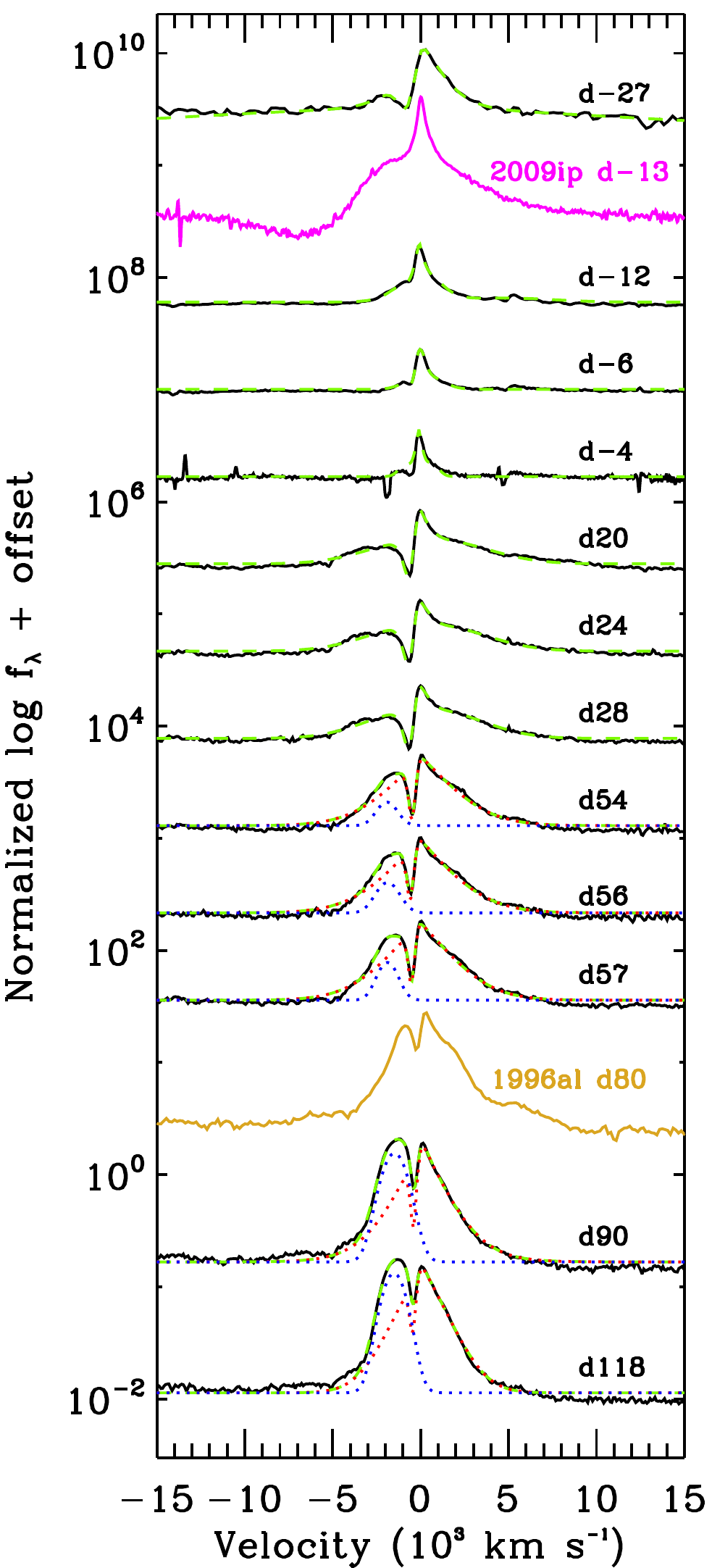}
	\caption{Our full spectral series centred on the H$\alpha$ line and in velocity space.  The first two spectral epochs are well fit by a Voigt profile with a P-Cygni absorption component (green dashed lines for day $-27$ to day $28$).  However, epochs after day $54$ exhibit significant blueshifted emission that appears asymmetric and cannot be fit by a single line profile with P-Cygni absorption.  We fit these line profiles with the same Voigt profile having P-Cygni absorption (red dotted line) plus an additional Gaussian component (blue dotted line).  The combined spectra (green dashed lines) are overlaid.  For comparison, we show the H$\alpha$ profiles of SN~2009ip-12B from before optical maximum \citep[day $-13$ from][]{childress+16} as well as SN~1996al from well after maximum light \citep[day $80$ from][]{benetti+16}.}\label{fig:halpha}
\end{figure}

We plot the evolution of the H$\alpha$ profile of Gaia16cfr in \autoref{fig:halpha}.  The pre-maximum epochs exhibit a P-Cygni profile, indicating that Gaia16cfr had a low-velocity expanding photosphere, likely from an optically-thick shell of CSM containing hydrogen.  In these two epochs, we fit the H$\alpha$ profile using a simple Voigt profile with Lorentzian and Gaussian components and blueshifted Gaussian absorption to track the P-Cygni absorption.  The Lorentzian FWHM is roughly $250~\text{km~s}^{-1}$ in the day $-27$ epoch, tracing the unshocked but radiatively excited CSM.  We interpret this velocity as the pre-outburst wind speed, which suggests that the relatively compact shell of CSM must have been ejected recently (as we discussed in \autoref{sec:optical-lc}).  The Gaussian component of the H$\alpha$ profile has a FWHM of $1700~\text{km~s}^{-1}$, and likely tracks shocked material swept up by the outburst ejecta or material entrained in the outburst.  When the outburst itself is still relatively young and optically thick, this broad H$\alpha$ line ought to trace the fastest, outer ejecta.  The fact that this velocity is relatively slow compared to that of core-collapse SNe suggests that there was little energy in the outburst and the $\sim10^{49}$~erg of radiative energy we calculated in \autoref{sec:optical-lc} may be close (e.g., within a factor of a few) to the total outburst energy.

In the day $-12$ epoch, the line profile is dominated by a strong Lorentzian component, although the FWHM of this line is only $1500~\text{km~s}^{-1}$.  This is curious, as the ejecta in the first epoch of observations exhibited a broader line width. In the optically-thick phase of ejecta-CSM interactions, line widths are dominated by electron scattering. This model predicts that photons are trapped in the ionised region of CSM behind the forward shock and diffuse outward via multiple scatterings \citep[see, e.g.,][]{chugai+01,smith+10b}. Therefore, we would normally predict broader H$\alpha$ line widths than those associated with normal outburst kinematics alone. But the day $-12$ profile is clearly narrower than on day $-27$ (\autoref{fig:halpha}). It is unlikely that the lower line velocity is caused by deceleration of the ejecta due to mass loading of CSM overrun by the forward shock, as we predict the ejecta have only recently encountered the inner shell of CSM during the day $-12$ epoch and the total mass of CSM is small (as discussed in \autoref{sec:dust}).

Instead, it is possible that as the ejecta first encounter the dusty shell of CSM, the photosphere of Gaia16cfr occurs {\it outside} of the forward shock as was predicted for the day 36 spectrum of SN~2006gy \citep[in][]{smith+10b}.  The extremely hot spectrum and anomalously narrow H$\alpha$ line profiles suggest that the ionisation front could diffuse outward through unshocked CSM.  This observation is supported by the fact that the Balmer decrement is high during this epoch (H$\alpha$/H$\beta\approx5.2$), again matching the physical scenario proposed for SN~2006gy \citep{smith+07}.  Rather than tracking H$\alpha$-emitting features from the shocked region, the thermal continuum emission discussed in \autoref{sec:continuum} is hot enough that we are only seeing into opaque CSM beyond the forward shock but excited by X-ray/UV radiation from the interior.  Thus, the H$\alpha$ line widths trace the $\sim250~\text{km~s}^{-1}$ CSM with additional Lorentzian broadening produced by electron scattering.

One of the most striking features in the spectral evolution of Gaia16cfr is the double-peaked H$\alpha$ profile that arises as early as day $20$ in the post-maximum spectrum.  This profile consists of broad, redshifted Lorentzian wings and blueshifted emission between $-2000$ and $-1000~\text{km~s}^{-1}$.  One might interpret the overall H$\alpha$ emission component as a single, broad profile with P-Cygni absorption near $-450~\text{km~s}^{-1}$, but the emission is clearly asymmetric as it becomes stronger in the day $54$ epoch, with a much broader Lorentzian wing on the red side than on the blue side.  Moreover, the evolution to late times suggests that most of the change comes from a blueshifted component in emission, possibly because the redshifted emission on the far side of the homologously expanding outburst is absorbed.

Therefore, we fit the overall profile with the same Voigt profile having P-Cygni absorption as above plus an added Gaussian component to match the blueshifted emission to all epochs past day $54$, as we demonstrate in \autoref{fig:halpha}.  This fit produces an excellent match to the overall profile, with the added Gaussian profile as the only difference from the earlier epochs. The Lorentzian profile is typical of CSM-interacting outbursts, with a broad FWHM ($3900$ and $2400~\text{km~s}^{-1}$) centred near zero velocity.  The added P-Cygni component is largely unchanged from the earlier epochs and is typically centred around $-450~\text{km~s}^{-1}$ with a FWHM of $500~\text{km~s}^{-1}$, indicating that the outburst is still expanding inside of an optically-thick region and may exhibit further CSM interaction as the ejecta evolve.

The added Gaussian emission exhibits the most dramatic evolution between day $24$ and day $118$.  It is centred between $-1900~\text{km~s}^{-1}$ and $-1500~\text{km~s}^{-1}$ throughout the evolution of this blueshifted feature.  The line also shifts to redder velocities over time, suggesting that the emission mechanism powering the overall blueshifted profile is becoming more optically thin over time and we are seeing deeper into the emission profile to slower-moving material.  This interpretation agrees with our observations of the [Ca\II] and Ca\II\ IR triplet emission, which suggest that we are seeing deeper inside of the transient to the unshocked ejecta.  At the same time, the profile becomes broader (FWHM = $1300~\text{km~s}^{-1}$ to $1600~\text{km~s}^{-1}$), perhaps because a larger fraction of high-velocity material is being uncovered over time.

Overall, the double-peaked feature strongly resembles that of SNe~IIn such as SN~1996al and 1996L \citep[we plot a spectrum of SN~1996al in \autoref{fig:halpha};][]{benetti+99,benetti+16} as well as SN~2015bh \citep{elias-rosa+16,thone+17} and spectra of UGC2773-OT around $15$--$34$~days after maximum light \citep{foley+11}. In all of these cases, the double-peaked line structure was interpreted as an imprint of a shocked inner shell of ejecta that arises as the outer CSM becomes optically thin \citep[see, e.g., the model in Figure~13 of][]{thone+17}.  The lack of a corresponding redshifted component is interpreted as absorption of the high-velocity material from dusty CSM along the line of sight to the far side of the interaction.

This H$\alpha$ velocity structure may be a generic feature of relatively low-energy explosions inside of a low-mass (for SNe~IIn) but compact shell of CSM.  \citet{benetti+16} identified SN~1996al as a $1.6\times10^{50}~\text{erg}$ explosion with $\sim1$~M$_{\odot}$ of ejecta expanding into $0.1$--$0.2$~M$_{\odot}$ of CSM.  Gaia16cfr likely had similar explosion properties, although this does not necessarily imply that Gaia16cfr was a core-collapse SN or that SN~1996al was the nonterminal explosion of a massive star.  Does a continuum exist between the most luminous objects identified as SN impostors and low-energy SNe~IIn, or are these transients physically distinct?  Continuous spectroscopic follow-up observations to late times is critical, as the H$\alpha$ profile may reveal the return to a quiescent LBV-like phase and suggest that the star is still bound.

\section{The Nature of Gaia16cfr and Other Luminous SN Impostors}\label{sec:comp}

From the pre-outburst and post-outburst data, we have assembled a picture of the Gaia16cfr progenitor system and its circumstellar environment.  Comparing these features to those of luminous SN impostors such as SN~2009ip-12B and SN~2015bh, we find the following.

\begin{enumerate}
	
\item[(1)] The optical SED of the Gaia16cfr progenitor source is consistent with an F8~I star, implying the progenitor star had a mass of $18$~M$_{\odot}$. However, the progenitor system was likely obscured by significant CSM extinction, and its implied luminosity, temperature, and mass must be treated as lower limits.  The SN~2009ip progenitor star was likely more luminous, blue, and with a much larger initial mass, perhaps $50$--$80$~M$_{\odot}$ \citep[as inferred from its luminosity of $\log(L/L_{\odot})=5.9$;][]{smith+10,foley+11}.  The SN~2015bh progenitor star was luminous, blue, and highly variable, although its exact mass is poorly constrained \citep{elias-rosa+16,thone+17}.

\item[(2)] Pre-outburst observations of Gaia16cfr from \hst\ in 2016 and from \spitzer\ in 2003 all suggest that the progenitor system had a significant IR excess from a relatively compact, dusty shell.  The dust mass in the immediate environment of the progenitor system is small ($4.2\times10^{-6}$~M$_{\odot}$), but the long baseline throughout these pre-outburst data suggests we tracked the source through multiple phases of its evolution.  It is possible that we observed multiple dust shells throughout this period and the progenitor source was episodically ejecting material at $\sim5\times10^{-4}$~M$_{\odot}~\text{yr}^{-1}$ over a decade before its major outburst.  Near-IR spectroscopy after the SN~2009ip-12B event suggests that this star was evolving into a $4\times10^{-7}$~M$_{\odot}$ shell of dust at a minimum radius of 120~AU \citep{smith+13}, closely matching the properties we found around Gaia16cfr.  SN~2015bh exhibited a small IR excess in pre-outburst data, although this emission was not variable until 180~days before the outburst \citep{thone+17}.

\item[(3)] The Gaia16cfr prognenitor source exhibited $1$--$2$~mag variability on timescales of weeks less than a year before outburst.  Given that the optical photospheric radius is consistent with that of a typical supergiant star, the progenitor system is consistent with exhibiting variability on the dynamical timescale of an F8~I star or from an optically-thick wind outside of the progenitor source.  Similar variability was seen before the SN~2009ip-12B event (the ``2011 eruptions'') with approximately the same magnitude and timescale roughly a year before the SN~2009ip-12A event \citep{fraser+13,pastorello+13,graham+14}, as well as in SN~2015bh \citep{thone+17}.

\item[(4)] The optical light curve of the Gaia16cfr outburst is remarkably similar to that of SN~2009ip-12B and SN~2015bh, especially given that all of these objects likely exhibited a precursor outburst followed almost immediately by a sharp rise to maximum light \citep[\autoref{fig:lc}, \autoref{fig:zoomed}, and][]{graham+14,thone+17}. The peak bolometric luminosity of Gaia16cfr was $-18.3$~mag and the decline time was initially $0.05$~mag~day$^{-1}$, almost exactly matching the characteristics of SN~2009ip-12B.  Also, like SN~2009ip-12B, Gaia16cfr exhibited a plateau in its light curve roughly 60~days after peak luminosity.  These characteristics are consistent with the interaction between ejecta from an outburst and a compact shell of CSM.  The later plateau suggests the CSM is structured beyond the main dust shell, possibly from previous mass ejections.  Furthermore, the timescale of interaction between ejecta and the main shell of CSM indicates that the dust observed in 2003 \spitzer\ data cannot be associated with the main dust shell.  These data all strongly imply that the progenitor system underwent episodic mass ejections before its major outburst in Dec. 2016.  The total integrated optical luminosity is $\sim 10^{49}~\text{erg}$, which is comparable to that of SN~2009ip-12B \citep[$3\times10^{49}$~erg in][]{graham+14}.

\item[(5)] The forward-shock velocity traced by the radius of the optical photosphere is $7500~\text{km~s}^{-1}$, while the velocity of the ejecta traced by the early-time FWHM of the Gaussian H$\alpha$ profile is about $1700~\text{km~s}^{-1}$.  SN~2009ip-12B exhibited $8000~\text{km~s}^{-1}$ line widths initially, which evolved to much faster velocities ($>10,000~\text{km~s}^{-1}$) in the post-maximum phase \citep{mauerhan+13,pastorello+13,graham+14}.  \citet{ofek+16} found that SN~2015bh exhibited ejecta velocities up to $15,000~\text{km~s}^{-1}$ in its nonterminal 2013 outburst.  These findings imply that massive-star outbursts can eject material up to extremely high velocities, but do not necessarily imply that the bulk of the ejecta are accelerated to these velocities or that this is a signature of a core-collapse SN.

\item[(6)] Gaia16cfr exhibited strong Ca\II\ IR triplet emission that broadened significantly at late times with little or no [Ca\II] emission, implying that most of the inner, unshocked ejecta were mostly obscured by optically-thick CSM.  In addition, Gaia16cfr had an extremely hot thermal continuum roughly 12~days before $r$-band maximum.  Combined with anomalously narrow H$\alpha$ features compared to the broader Gaussian features from early times, these data suggest that CSM exterior to the forward shock formed a photosphere when it was ionised by strong X-ray/UV radiation from the shocked region.  SN~2009ip-12B exhibited little Ca\II\ IR triplet or [Ca\II] emission until late times \citep{graham+14}.  SN~2015bh exhibited both the Ca\II\ IR triplet and [Ca\II] emission at $>100$~days after optical maximum, which imply an ongoing CSM interaction but one that rapidly became optically thin \citep{elias-rosa+16}.

\item[(7)] The H$\alpha$ profile of Gaia16cfr was highly structured as it declined past maximum light.  In addition to a typical P-Cygni profile, the overall line profile exhibited a strong blueshifted emission feature that became stronger over time.  We interpret this line profile as an indication that the outer CSM is becoming optically thin and revealing high-velocity ejecta from the outburst itself \citep[as in SN~2015bh;][]{thone+17}.  SN~2009ip-12B exhibited a broad H$\alpha$ profile and a narrow Lorentzian profile with FWHM $=500$--$1000~\text{km~s}^{-1}$ \citep{mauerhan+13}.  As the outburst evolved to late times, the broad component increased in width to $\sim15,000~\text{km~s}^{-1}$.  SN~2015bh was spectroscopically similar to Gaia16cfr in the post-maximum phase, with the same double-peaked H$\alpha$ structure \citep{elias-rosa+16,thone+17}.  We find that all of these events had a similar structure to the double-peaked SNe~IIn 1996L and 1996al \citep{benetti+99,benetti+16}, implying that some low-energy SNe~IIn share a similar CSM structure to luminous SN impostors. 

\end{enumerate}

The comparisons between Gaia16cfr, SN~2009ip-12B, and SN~2015bh are particularly intriguing, especially in light of the interpretation that some, all, or none of these events may be core-collapse SNe.  It is possible that these similarities can be explained in large part by their circumstellar environments, as they appear to have exploded in relatively dense but compact dust shells.  As their optical light curves are largely dominated by emission from the CSM interaction region, the shape and peak luminosity of their light curves might be attributed to similar CSM configurations.  Precursor variability implies that episodic mass ejections produced a dense, structured circumstellar environment, but it is not clear why the progenitor systems exhibited significant variability at roughly the same epoch before their outbursts, or what connections exist in the physical mechanism responsible for the outbursts.  Does this combination of precursor variability and major outbursts occur in stars with a range of masses \citep[e.g., from IRC+10420 to $\eta$ Car;][]{smith+04}, or do the similarities between SN~2009ip, SN~2015bh, and Gaia16cfr imply that the latter came from a massive but heavily obscured progenitor star?  A wider sample of progenitor stars from SN impostors and SNe~IIn is necessary to answer this question, as well as follow-up observations of Gaia16cfr to confirm the final fate of the progenitor star.

Perhaps the most striking difference is the order-of-magnitude discrepancy in their H$\alpha$ FWHM values.  \citet{smith+14} argue that the true ejecta mass of SN~2009ip-12B was likely $4$--$6$~M$_{\odot}$ with an average velocity of $4500~\text{km~s}^{-1}$ (i.e., an explosion energy of $10^{51}~\text{erg}$), and therefore a core-collapse SN was required to provide enough energy for the outburst.  It is unclear whether this ejecta mass and explosion energy apply for the bulk of the ejecta in SN~2009ip-12B, but the optical spectroscopy unambiguously demonstrates a broad (FWHM $=8000~\text{km~s}^{-1}$) component combined with high-velocity blueshifted absorption (\autoref{fig:spectra}), implying that some fraction of the material was accelerated to $\sim10,000~\text{km~s}^{-1}$.  Similar spectroscopic features were observed in spectra of SN~2015bh as early as 2~yr before outburst \citep{ofek+16,thone+17}, implying that SN~2015bh had a nonterminal outburst and accelerated ejecta to velocities as high as 15,000~km~s$^{-1}$.  Clearly, the presence of high-velocity ejecta does not necessarily imply that these events were core-collapse SNe, although they may indicate that there was significant asymmetry in the outburst mechanism or simply high-velocity knots of ejecta.

Comparisons between Gaia16cfr, SN~2015bh, and low-energy SNe~IIn such as SNe~1996L and 1996al may offer a method for distinguishing these events by their physical mechanisms.  Although it is generally accepted that the double-peaked H$\alpha$ profile observed from these objects is simply an imprint of an explosion inside of a low-mass and compact shell of CSM, it is possible that a single physical mechanism can explain all of these objects.  Deep, high-resolution imaging and spectroscopy at late times will be critical for drawing a self-consistent explanation between this configuration and an explosion model.

\section{CONCLUSIONS}

In this work, we used pre-outburst \hst\ and \spitzer\ data as well as post-outburst, ground-based photometry and spectroscopy to constrain the pre-outburst configuration of Gaia16cfr, its environment, and its outburst properties.  

The progenitor source we detected in pre-outburst images is consistent with a $\sim18$~M$_{\odot}$ progenitor star, although a significant IR excess in the \hst/$F160W$ and \spitzer\ bands suggest that this source is significant reddened by circumstellar dust.  We accurately modeled the dust configuration from 13~yr prior to outburst as a $690$~K shell at $120$~AU from the progenitor star with $4\times10^{-6}$~M$_{\odot}$ of dust.  Given typical CSM velocities in the environment of Gaia16cfr, it is unlikely that this shell was associated with the immediate circumstellar environment of Gaia16cfr at the time of outburst, although this detection implies that the progenitor source was periodically ejecting shells of material within years to decades of its major outburst.

\hst\ photometry within a year of the major outburst of Gaia16cfr indicates that it was ``flickering'' with a period of $10$--$30$~days and peak-to-peak changes of more than $1$~mag.  This flickering is remiscent of the 2011 outbursts of SN~2009ip \citep{pastorello+13,graham+14} and periodic variability preceding SN~1954J and SN~2015bh \citep{tammann+68,thone+17}.  Combined with the H$\alpha$ luminosity observed from the progenitor source in 2006, we interpret this variability as periodic ejections of material in an optically-thick wind, similar to LBVs during their ``maximum'' outbursting phase \citep[see, e.g., Figure 4 of][]{lamers95}.

Immediately before, during, and after its rise to maximum brightness, Gaia16cfr exhibited almost the exact same light curve as SN~2009ip-12B, with precursor variability, a sharp rise to peak, bolometric peak magnitude near $-18$~mag, a gradual decline, and an eventual plateau roughly 60~days after optical maximum.  This similarity is likely caused by the similar circumstellar environments of these two events and the ejection of a high-velocity shell of material that encountered shells of CSM over time.

Spectroscopically, Gaia16cfr is similar to SN~2015bh and low-energy SNe~IIn such as SNe~1996al and 1996L \citep{benetti+99,benetti+16,elias-rosa+16}.  In Gaia16cfr, the evolution of Ca emission and absorption, thermal continuum, and H$\alpha$ all indicate that the high-velocity ejecta encountered a shell of CSM that became optically thin over time, revealing inner layers of ejecta.  The comparison to low-energy SNe~IIn such as SNe~1996al and 1996L suggests that a continuum exists between this class of objects and SN impostors, but the nature of this relation and the connection to the relevant physical mechanisms is still ambiguous.

Continued monitoring of Gaia16cfr, especially deep, high-resolution follow-up optical imaging and spectroscopy, will be critical to discovering the physical mechanism powering this object and other SN impostors.  While the possibility that all of these objects are core-collapse SNe is still open to debate, a late-time detection of a surviving progenitor that resembles a quiescent LBV is perhaps the most promising method for finally resolving this issue.
	
\smallskip\smallskip\smallskip\smallskip
\noindent {\bf ACKNOWLEDGMENTS}
\smallskip
\footnotesize

We would like to thank Tyler Takaro, Draco Reed, and Rajdipa Chowdhury for assistance with SOAR data acquisition, as well as Nahir Mu\~{n}oz Elgueta, Michael Foley, Jorge Anais, Natalie Ulloa, and Abdo Campillay for their help with Swope observations.  We are grateful to our \hst\ program coordinator, Amber Armstrong, for help executing GO-13646.

The UCSC group is supported in part by NSF grant AST-1518052 and from fellowships from the Alfred P.\ Sloan Foundation and the David and Lucile Packard Foundation to R.J.F. The work of A.V.F. was conducted in part at the Aspen Center for Physics, which is supported by NSF grant PHY-1607611; he thanks the Center for its hospitality during the neutron stars workshop in June and July 2017. A.V.F. is grateful for financial assistance from the TABASGO Foundation, the Christopher R. Redlich Fund, the Miller Institute for Basic Research in Science (U.C. Berkeley), and \hst\ grants GO-13646 and AR-14295 from the Space Telescope Science Institute (STScI), which is operated by AURA under NASA contract NAS 5-26555.

This work includes data obtained with the Swope Telescope at Las Campanas Observatory, Chile, as part of the Swope Time Domain Key Project (PI Piro, Co-PIs Shappee, Drout, Madore, Phillips, Foley, and Hsiao).
This work is based in part on observations collected at the European Organisation for Astronomical Research in the Southern Hemisphere, Chile as part of PESSTO (the Public ESO Spectroscopic Survey for Transient Objects Survey) ESO program 188.D-3003, 191.D-0935, and 197.D-1075.

This work is based in part on observations made with the {\it Spitzer Space Telescope}, which is operated by the Jet Propulsion Laboratory, California Institute of Technology, under a contract with NASA.  The {\it Hubble Space Telescope} (\hst) is operated by NASA/ESA. The \hst\ data used in this manuscript come from programs GO-10803, GO-13646, and GO-14645 (PIs Smartt, Foley, and Van Dyk, respectively). Some of our analysis is based on data obtained from the \hst\ archive operated by STScI. Our analysis is based in part on observations obtained at the Southern Astrophysical Research (SOAR) telescope, which is a joint project of the Minist\'{e}rio da Ci\^{e}ncia, Tecnologia, e Inova\c{c}\~{a}o (MCTI) da Rep\'{u}blica Federativa do Brasil, the U.S. National Optical Astronomy Observatory (NOAO), the University of North Carolina at Chapel Hill (UNC), and Michigan State University (MSU).

\textit{Facilities}: \hst\ (ACS/WFC3), \spitzer\ (IRAC), Swope (LCO), SOAR (Goodman)

\bibliography{gaia16cfr}

\end{document}